\documentclass[english]{article}

\usepackage[T1]{fontenc}
\usepackage{amssymb}
\usepackage{amsmath}
\usepackage{amsfonts}
\usepackage{graphicx}
\usepackage{latexsym}
\usepackage{bm}
\usepackage{tensor}
\usepackage{hyperref}

\begin{document}
	
\title{Spectra of electronic excitations in graphene near Coulomb impurities}
	
\author{A.~I.~Breev$^{1}$\thanks{breev@mail.tsu.ru}, R.~Ferreira$^{2}$\thanks{rafaelufpi@gmail.com}, D.~M.~Gitman$^{1,2,3}$\thanks{gitman@if.usp.br}, and B.~L.~Voronov$^{3}$\thanks{voronov@lpi.ru} \\
{\normalsize $^{1}$ Department of Physics, Tomsk State University, Tomsk
634050, Russia;}\\
{\normalsize $^{2}$ Institute of Physics, University of S\~{a}o Paulo, CEP
05508-090, S\~{a}o Paulo, SP, Brazil;}\\
{\normalsize $^{3}$ P.N. Lebedev Physical Institute, 53 Leninsky prospekt,
119991 Moscow, Russia;}}

\maketitle

\begin{abstract}
We study the problem of the electron excitation spectrum in the presence of
point-like and regularized Coulomb impurities in gapped graphene. To this
end, we use the Dirac model and in the point-like case theory of
self-adjoint extensions of symmetric operators. In the point-like case, we
construct a family of self-adjoint Hamiltonians describing the excitations
for any charge of an impurity. Spectra and (generalized) eigenfunctions for
all such Hamiltonians are found. Then, we consider the spectral problem in
the case of a regularized Coulomb potential of impurities for a special
regularization. We study exact equations for charges of impurities that may
generate bound states with energy that coincides with the upper boundary of
the negative branch of the continuous spectrum (supercritical charges) and
calculate these charges.
\end{abstract}


\section{Introduction\label{S1}}



It is well known that low-energy electronic excitations in the graphene
monolayer in the presence of an external electromagnetic field can be
described by the Dirac model with the corresponding background \cite{Sem84},
namely, by a $2+1$ quantized Dirac field in such a background. In fact, in
the framework of the latter model, it is more correct to speak about some
quasiparticles, which are chiral Dirac fermions in $2+1$ dimensions. Taking
into account that dispersion surfaces are the so-called Dirac cones, the gap
between the upper and lower branches in the corresponding Dirac particle
spectra is very small and charge carriers are massless, we can expect that
Schwinger mechanism of vacuum instability is essential already in laboratory
conditions, almost any electric-like external field is supercritical\footnote{
Note also that the standard QED vacuum in (3+1) dimensions becomes unstable
due to the Coulomb attraction between electron and positron above a critical
value of the fine-structure constant \cite{Goldstein,SU}, $\alpha_{cr}=\pi/8$ 
or, with its genuine value of $\alpha =1/137$, but if an external magnetic 
field above $10^{42}$ G is imposed \cite{SU}.} and the particle creation effect turns out 
to be dominant (under certain conditions) in various quantum processes in the 
external electric-like field in the graphene. Application of QED with strong 
field and unstable vacuum to describing quantum processes in graphene requires 
nonperturbative methods. In particular, the conductivity of graphene, 
especially in the so-called nonlinear regime, was studied in Ref. 
\cite{Git2012} using such methods. The production of electron-hole pairs was 
recently observed in graphene while investigating a behavior of its current-voltage 
characteristics \cite{Vand2010}.

It is known that one of the graphene property is its great sensitivity to
impurities and defects, which is associated with the vanishing density of
states of current carriers. The transport of electrons in the presence of
charged impurities such as Coulomb centers becomes one of the most important
topics relating to achieving maximum carrier mobility in graphene
\cite{Nomura}. Because the Coulomb law remains unchanged in undoped graphene and
is approximately retained for small and moderate doping, the scattering
processes are essentially controlled by an unscreened Coulomb potential,
unlike ordinary metals where the screening is complete \cite{Kotov2012}.

A potential induced by substrate can break the symmetry of the honeycomb
lattice and generate a gap in the graphene electron spectrum. The gap then
suggests that the motion of electrons can be described by the $2+1$ massive
Dirac equations. The problem of the electron spectra of gapped graphene with
Coulomb impurities was considered in \cite{Pereira2, Zhu, Godb, Godb2}.

Relativistic scattering theory for $2+1$ Dirac electrons in graphene by
Coulomb impurities uses solutions of spectral problem for a Dirac
Hamiltonian in $2+1$ dimensions with the corresponding point-like or
regularized three-dimensional Coulomb potential. The corresponding spectral
problem in $3+1$ dimensions was discussed in detail in
Refs. \cite{GVT, book2012, PhScr2013, Kyl} for the point-like Coulomb
potential and in Refs. \cite{Kyl, VorGitLevFer2016} for a regularized Coulomb
potential. For the point-like Coulomb potential a consistent treatment of this
problem depends essentially on a correct definition of the Dirac Hamiltonian as a
self-adjoint (s.a. in what follows) operator in an appropriate Hilbert
space. Whereas in $3+1$ dimensions peculiarities of such a definition
apppear only for nuclei with large $Z>119$, which do not exist in laboratory
conditions, for Coulomb impurities in the graphene this threshold is much
less due to properties of Dirac excitations in the graphene.

In this paper, we consider the problems of correctly defining the Dirac
Hamiltonian for quasiparticles in gapped graphene in the presence of a
Coulomb impurity of a charge $Z$ as the s.a. operator and of its spectral
analysis. We present a rigorous consideration of all aspects of these
problems based on theory of s.a. extensions of symmetric operators \cite{book2012}.
We also study the spectral problem in the case of a regularized
Coulomb field of impurities, which allows an exact analytical formulation.

The paper is organized as follows. We begin with introducing a basic
information and equations explaining the setting of the problem, Section \ref{S2}.
In Section \ref{S3}, we describe a mathematically rigorous procedure
for reducing the problem of constructing s.a. rotationally invariant Dirac
Hamiltonian in the full Hilbert space to the problem of constructing s.a.
one-dimensional partial radial Hamiltonians with certain angular momenta.
The further consideration is divided into two parts. In the first part
(Sections \ref{S4}-- \ref{S5.3}), we consider the point-like Coulomb field
of impurities in gapped graphene. In Section \ref{S4}, we construct a
general solution of radial equations, as well as some particular solutions
of these equations used in what follows. Section \ref{S5} is devoted to
constructing s.a. partial radial Hamiltonians with arbitrary admissible
angular momentum $j$ and to solving the corresponding spectral problems in
four different regions of the upper ($j$, $Z$) half-plane, the regions differ
by a character .of s.a. boundary conditions at the origin specifying partial
radial Hamiltonians. Section \ref{S5.3} is devoted to desribing pecularities
of the total Hamiltonian\ in dependence of a charge $Z$. The second part of
the paper (Sections \ref{r}--\ref{R3}) deals with a regularized Coulomb
field of impurities in the gapped graphene. In the case of a specially
regularized Coulomb field, we deduce the exact equations which allow finding
the point spectrum (located in the semi-interval $[-m,m)$) together with the
corresponding bound states,\ Section \ref{r}. In the Section \ref{R2}, we
present the exact equations for charges that give bound states with energy
$E=-m$ that coincides with the upper boundary of the negative branch of the
continuous spectrum. In the Section \ref{R3}, we discuss the concept of
supercritical impurity charge. Section \ref{S5.4d} is devoted to a
discussion of our results in the both parts of the paper.

\section{Dirac equation in $2+1$ dimensions with point-like Coulomb field
\label{S2}}

Here we will consider the spectrum of quasiparticle excitations in the
presence of a single Coulomb impurity of charge $Ze$ placed in the middle of
the hexagon of the graphene lattice. More specifically, we restrict
ourselves by neighborhoods of the two so called Dirac points which are the
regions of the quasiparticle spectrum most sensitive to an
influence of the impurity. In addition, we suppose that the impurity is a
fully stripped ion, so that $Z$ is an atomic number.

We introduce a Cartesian coordinate system with $Ox$ and $Oy$ axes lying in
the graphene plane and the $Oz$ axis normal to the plane. The impurity is
situated at the origin. The Coulomb field is weakened by the polarization of
the crystal at distances greater than the interatomic ones. To take this
effect into account, a macroscopic permittivity $\epsilon$ (which is also
called an effective dielectric permittivity) must be introduced into the
Coulomb potential, which reads
\begin{equation*}
V(\rho )=-\frac{Ze^{2}}{\epsilon }\left( \frac{1}{\rho }\right) ,\quad
\rho=\left\vert \mathbf{r}\right\vert =\sqrt{x^{2}+y^{2}},
\end{equation*}
where $\mathbf{r=(}x,y)$ is a two-dimensional vector (in the graphene plane
$z=0$).

Quasiparticles in graphene live in two-dimensional space, whereas the
Coulomb field created by an impurity lives in three-dimensional space, and
therefore the Coulomb potential in the graphene plane $z=0$ behaves like
$\rho^{-1}$. If we were to consider the electric field created by an
impurity of charge $q$ that lives in two-dimensional space, the electric
potential $\Phi$ would be $\Phi=q/(2\pi \epsilon)\log (1/\rho )$ as
it follows from the Poisson equation
$\Delta \Phi = -(q/\epsilon)\delta(\mathbf{r})$ in two-dimensional space.

In the case that the shielding is due to electrons in graphene, the RPA
method gives $\epsilon \approx 5$ (see Ref. \cite{GonEp}). If we completely
neglect effects of a polarization in the graphene, but the graphene is on a
$\mathrm{SiO}_{2}$ substrate, then $\epsilon \approx 2.5$. (see also Ref. \cite{Ando2006}).

We note that potentials induced by the substrate can violate the symmetry of
the lattice and create gaps in the electronic spectrum: the gaps between the
conduction band and the valence band. By choosing a substrate, it is
possible to adjust the effective mass of charge carriers and investigate
phenomena absent in the massless case. In this paper, we assume that charge
carriers in graphene can have some effective mass $m_{\ast }>0$.

To distinguish the two different Dirac points in the graphene, we introduce
a parameter, or an index, $s=\pm 1$, which we will call isospin in the
following because of a certain analogy with the latter. We let $\mathbf{K}_{s}$
denote the Dirac points, their coordinates in the Brillouin zone are
chosen as $\mathbf{K}_{s}=(4\pi s/(3a),0)$, where $a=2.46\mathring{A}$
is the lattice constant.

As we said above, we restrict our consideration to the neighborhoods of the
two Dirac points, which means that we restrict ourselves to low-energy
quasiparticle excitations and low-energy quasiparticle transitions. In this
approximation, the complete Hilbert space $\mathfrak{H}_{\mathrm{tot}}$ of
quantum states of an quasiparticle in graphene is a direct orthogonal sum of
two Hilbert spaces $\mathfrak{H}_{s}$, $s=\pm 1$, each of which is related
to the corresponding Dirac point $\mathbf{K}_{s}$, the both Hilbert spaces
$\mathfrak{H}_{s}$ are Hilbert spaces of two-dimensional doublets, so that
$\mathfrak{H}_{\mathrm{tot}}=\mathfrak{H}_{1}\oplus \mathfrak{H}_{-1}$,
$\mathfrak{H}_{1}=\mathfrak{H}_{-1}=\mathfrak{H=}\,L^{2}(\mathbb{R}^{2})\oplus
L^{2}(\mathbb{R}^{2})$.

Usually, in the graphene, intervalley processes are not taken into account
referring to the long-range nature of the Coulomb field. Formally, this
means that transitions between the Hilbert spaces $\mathfrak{H}_{s}$ are not
considered, such that the total quasiparticle Hamiltonian $\hat{H}_{\mathrm{tot}}$
is a direct orthogonal sum of two Hamiltonians $\widehat{\mathcal{H}}_{s}$,
$s=\pm 1$, each acting in the respective Hilbert space $\mathfrak{H}_{s}$ and
can be treated independently.

In the framework of the tight-binding approximation and for low
quasiparticle energy, the stationary Schr\"{o}dinger equation for an
electron reduces to the following two independent equations for the
envelopes of Bloch functions in the neighborhood of each Dirac point
$\mathbf{K}_{s}$ \cite{Raza}:
\begin{equation}
\mathcal{\check{H}}_{s}\Psi _{s}=\mathcal{E}\Psi _{s},\quad s=\pm 1,
\label{1.2a}
\end{equation}
where wave functions $\Psi_{s}$ are doublets depending on $\mathbf{r}$,
$\Psi_{s}=\Psi_{s}(\mathbf{r})=\{\psi_{s\alpha }(\mathbf{r})$, $\alpha=1,2\}$,
whose components $\psi_{s\alpha}(\mathbf{r})$ are the envelopes of the
Bloch functions in two respective sublattices $A$ and $B$, and
$\mathcal{\check{H}}_{s}$ are the corresponding Dirac differential operations:
\begin{eqnarray}
&&\mathcal{\check{H}}_{s}=\hbar v_{F}\left(
-i\left[ s\sigma _{x}\partial_{x}+\sigma _{y}\partial_{y}\right] -\frac{g}{\rho}\right)
+\frac{\Delta_{dop}}{2}\sigma_{z},  \notag \\
&&g=\frac{1}{\hbar v_{F}}\frac{Ze^{2}}{\epsilon}=\alpha _{F}\frac{Z}{\epsilon}=
\alpha_{F} Z_{eff},\quad Z_{eff}=\frac{Z}{\epsilon},
\label{1.2b}
\end{eqnarray}
where $v_{F}\approx 10^{6}$ cm/s is the Fermi velocity,
$\Delta_{dop}=2m_{\ast} v_{F}^{2}$ is the energy gap,
$\alpha _{F}$ $=e^{2}/(\hbar v_{F})$ is the graphene
\textquotedblleft fine structure constant\textquotedblright,
and $\{\sigma_{x}$, $\sigma_{y}$, $\sigma_{z}\}$ are the Pauli matrices.

In what follows, we use the following notation:
$\check{H}_{s}=(\hbar\nu _{F})^{-1}\mathcal{\check{H}}_{s}$,
$E=(\hbar \nu _{F})^{-1}\mathcal{E}$ and $m=(2\hbar v_{F})^{-1}\Delta_{dop}$.
Note that the variable $E$ and the parameter $m$ are of dimension of the
inverse length. Equation (\ref{1.2a}) then reads
\begin{equation}
\check{H}_{s}\Psi_{s}(\mathbf{r})=E \Psi_{s}(\mathbf{r}),\quad s=\pm 1,
\label{1.3a}
\end{equation}
where the differential operations $\check{H}_{s}$ in the Cartesian and in
the polar coordinates $\rho$, $\phi$, ($x=\rho \cos \phi $, $y=\rho \sin \phi$)
have the respective forms:
\begin{align}
\nonumber
 \check{H}_{s}= &-i\left( s\sigma _{x}\partial _{x}+\sigma _{y}\partial_{y}\right)
 -\frac{g}{\rho }+m\sigma _{z}  \\
 = &-i(s\cos \phi \sigma _{x}+\sin \phi \sigma _{y})\frac{\partial }{\partial \rho }+\frac{i}{\rho }(s\sin \phi \sigma _{x}-\cos \phi \sigma _{y})
\frac{\partial }{\partial \phi } -\frac{g}{\rho }+m\sigma _{z}.
\label{1.3b}
\end{align}

To assign a specific meaning of a quantum-mechanical eigenvalue problem for
certain Hamiltonians $\hat{H}_{s}$ to Eqs. (\ref{1.3a}), we have to solve
the two problems. The first one is to define, or construct,
the Hamiltonians $\hat{H}_{s}$ as s.a. operators with certain
domains in the Hilbert space
$\mathfrak{H}=L^{2}(\mathbb{R}^{2})\oplus L^{2}(\mathbb{R}^{2})$
of doublet functions acting on their domains by the
respective differential operations $\check{H}_{s}$ (\ref{1.3b}) (for
brevity, we will say that the operators $\hat{H}_{s}$ are associated
with differential operations $\check{H}_{s}$). The second problem is to
perform the spectral analysis of the obtained Hamiltonians, i.e., to
evaluate their spectra and the corresponding (generally generalized)
eigenfunctions. In solving the both problems, we follow the way adopted in
Refs. \cite{GVT} in the case of the $3$-dimensional Dirac equation with
point-like and regularized Coulomb fields.

\section{Reduction to radial problem\label{S3}}

We begin with defining an initial symmetric operators
$\hat{H}_{s}^{\text{\emph{in}}}$ in the Hilbert space
$\mathfrak{H=}L^{2}(\mathbb{R}^{2})\oplus L^{2}(\mathbb{R}^{2})$ associated
with the respective differential operations $\check{H}_{s}$ (\ref{1.3b}).
Because coefficient functions of differential operations $\check{H}_{s}$ are
smooth outside the origin, we choose the space of smooth compactly supported
doublets for the domains $D(\hat{H}_{s}^{\text{\emph{in}}})$ of
$\hat{H}_{s}^{\text{\emph{in}}}$.

Thus, one can avoid troubles with a behavior of doublets at infinity. To
avoid troubles with the $1/\rho $ singularity of the Coulomb potential at
the origin, we additionally require that all doublets in
$D(\hat{H}_{s}^{\text{\emph{in}}})$ vanish in some neighborhood of the origin, specific to
each doublet. Note that the domains $D(\hat{H}_{s}^{\text{\emph{in}}})$
(which are the same for the both values $s$) are dense in $\mathfrak{H}$.
The operators $\hat{H}_{s}^{\text{\emph{in}}}$ are thus defined as
\begin{equation*}
\hat{H}_{s}^{\text{\emph{in}}}=\left\{
\begin{array}{l}
D(\hat{H}_{s}^{\text{\emph{in}}})=\left\{ \Psi (\mathbf{r}):
\psi _{\alpha }(\mathbf{r})\in C_{0}^{\infty }(\mathbb{R}^{2}\setminus \{0\})\right\}, \\
\hat{H}_{s}^{\text{\emph{in}}}\Psi (\mathbf{r})=\check{H}_{s}\Psi (\mathbf{r}).
\end{array}
\right. 
\end{equation*}
The symmetricity of $\hat{H}_{s}^{\text{\emph{in}}}$ is evident.

We construct s.a. Hamiltonians $\hat{H}_{s}$ as s.a. extensions of the
respective initial symmetric operators $\hat{H}_{s}^{\text{\emph{in}}}$, to
emphasize this circumstance, we introduce an additional index $\mathfrak{e}$
to $\hat{H}_{s}$, $\hat{H}_{s}\rightarrow \hat{H}_{s}^{\mathfrak{e}}$. There
exist different s.a. extensions of a given $\hat{H}_{s}^{\text{\emph{in}}}$,
such that the index $\mathfrak{e}$ will be replaced by a more informative
index in what follows.

We require that $\hat{H}_{s}^{\mathfrak{e}}$ be rotationally invariant as
well as the initial symmetric operators $\hat{H}_{s}^{\text{\emph{in}}}$
are. The meaning of this requirement is explained below.

There are the two different unitary representations $U_{s}$ of the
rotation group \textrm{Spin}(2) in $\mathfrak{H}$ which are connected with
the respective operators $\hat{H}_{s}^{\text{\emph{in}}}$. The generator
$\hat{J}_{s}$ of the representation $U_{s}$, conventionally called
the angular momentum operator (there are two of them), is a s.a. operator in
$\mathfrak{H}$ defined on absolutely continuous and periodic in
$\phi \in \lbrack 0,2\pi ]$ doublets and associated with the differential operation
$\check{J}_{s}=-i\partial /\partial \phi +s\sigma _{z}/2$. For each $s$, the
Hilbert space $\mathfrak{H}$ is represented as a direct orthogonal sum
\begin{equation}
\mathfrak{H=}\sum_{j}{}^{\oplus }\mathfrak{H}_{sj},\quad
j=\pm 1/2,\pm 3/2\dots,  \label{3.3}
\end{equation}
of subspaces $\mathfrak{H}_{sj}$ that are the eigenspaces of the
angular momentum operator $\hat{J}_{s}$ corresponding to all its eigenvalues
$j=\pm 1/2,\pm 3/2\dots$. The subspace $\mathfrak{H}_{sj}$ with given $s$,
$j$ consists of doublets $\Psi _{sj}$ of the form
\begin{equation}
\Psi _{sj}(\mathbf{r})=\frac{1}{\sqrt{2\pi \rho }}e^{ij\phi }
\begin{pmatrix}
e^{-is\phi /2}f(\rho ) \\
-ise^{+is\phi /2}g(\rho )
\end{pmatrix}
\in \mathfrak{H}_{sj},  \label{3.4}
\end{equation}
they are the eigenfunctions of $\hat{J}_{s}$ with the eigenvalue $j$,
$\hat{J}_{s}\Psi_{sj}(\mathbf{r})=\check{J}_{s}\Psi_{sj}(\mathbf{r})=
j\Psi _{sj}(\mathbf{r})$. We note that spectra of both operators $\hat{J}_{-1}$ and
$\hat{J}_{1}$ are the same. The functions $f(\rho)$ and $g(\rho)$ are
called the radial functions. In the physical language, decompositions (\ref{3.3})
and (\ref{3.4}) correspond to the expansion of doublets $\Psi(\mathbf{r})\in \mathfrak{H}$
in terms of eigenfunctions of the two different angular momentum operators
$\hat{J}_{-1}$ and $\hat{J}_{1}$.

The following remark is very useful in our constructions. Let
$\mathbb{L}^{2}(\mathbb{R}_{+})$ be the Hilbert space of radial doublets,
\begin{equation*}
F(\rho )=\left(
\begin{array}{c}
f(\rho ) \\
g(\rho )
\end{array}
\right) \in \mathbb{L}^{2}(\mathbb{R}_{+}),
\end{equation*}
with the scalar product
\begin{align*}
(F_{1},F_{2})=&\int\limits_{0}^{\infty }F_{1}^{~+}(\rho )F_{2}^{~}(\rho)d\rho
 =\int\limits_{0}^{\infty }\left[ \overline{f_{1}(\rho )}~f_{2}(\rho)+
 \overline{g_{1}(\rho )}~g_{2}(\rho )\right] d\rho,
\end{align*}
so that $\mathbb{L}^{2}(\mathbb{R}_{+})=L^{2}(\mathbb{R}_{+})\oplus L^{2}(\mathbb{R}_{+})$.
Then Eq. (\ref{3.4}) and the relation
\begin{equation*}
\left\Vert \Psi _{sj}\right\Vert ^{2}=\int\limits_{0}^{\infty }
\left[|f(\rho )|^{2}+|g(\rho )|^{2}\right] d\rho
\end{equation*}
show that each subspace $\mathfrak{H}_{sj}\subset \mathfrak{H}$ is
unitarily equivalent to $\mathbb{L}^{2}(\mathbb{R}_{+})$,
\begin{equation}
\Psi _{sj}(\mathbf{r})=V_{sj}F(\rho ),\quad
F(\rho )=V_{sj}^{-1}\Psi _{sj}(\mathbf{r}).  \label{3.9}
\end{equation}
If necessary, an explicit form of operators $V_{sj}$ and $V_{sj}^{-1}$ can
be easily written down.

The initial symmetric operators $\hat{H}_{s}^{\text{\emph{in}}}$ are
rotationally invariant. Namely, each $\hat{H}_{s}^{\text{\emph{in}}}$ is
invariant from the standpoint of the representation $U_{s}$ of the rotation
group. By definition, this means that each subspace $\mathfrak{H}_{sj}$ (the
eigenspace of the generator $\hat{J}_{s}$ with eigenvalue $j$) reduces the
operator $\hat{H}_{s}^{\text{\emph{in}}}$. In other words, the operator
$\hat{H}_{s}^{\text{\emph{in}}}$ commutes with orthoprojectors $P_{sj}$ on
the subspaces $\mathfrak{H}_{sj}$, see \cite{AkhGl81}. This means the
following. Let
$\Psi _{s}(\mathbf{r})=\sum_{j}\Psi _{sj}(\mathbf{r})\in D(\hat{H}_{s}^{\text{\emph{in}}})$.
Then
\begin{equation*}
\Psi _{sj}=P_{sj}\Psi_{s}\in D(\hat{H}_{s}^{\text{\emph{in}}}),\quad
\hat{H}_{s}^{\text{\emph{in}}}\Psi_{s}=\sum\limits_{j}
\hat{H}_{sj}^{\text{\emph{in}}}\Psi _{sj},
\end{equation*}
where $\hat{H}_{sj}^{\text{\emph{in}}}=P_{sj}\hat{H}_{s}^{\text{\emph{in}}}P_{sj}=
\hat{H}_{s}^{\text{\emph{in}}}P_{sj}$ are the so called parts of
$\hat{H}_{s}^{\text{\emph{in}}}$ lying in $\mathfrak{H}_{sj}$, their rule of
acting is given by the certain first-order differential operation in $\rho$,
which is easily evaluated, we will see it below. Thus, each initial
symmetric operator $\hat{H}_{s}^{\text{\emph{in}}}$ is a direct orthogonal
sum of its parts,
\begin{equation*}
\hat{H}_{s}^{\text{\emph{in}}}=\sum\limits_{j}{}^{\oplus}
\hat{H}_{sj}^{\text{\emph{in}}},
\end{equation*}
and a study of the rotationally invariant $\hat{H}_{s}^{\text{\emph{in}}}$
is reduced to a study of its parts $\hat{H}_{sj}^{\text{\emph{in}}}$. Note
that these facts are essentially based on the formal commutativity of the
differential operations $\check{H}_{s}$ and $\check{J}_{s}$,
$[\check{H}_{s},\check{J}_{s}]=0$.

We would like to make in passing a general note concerning a rather popular
understanding of quantum-mechanical symmetry in physical literature by the
above example of rotational symmetry. In physical literature, the statement
that the operators $\hat{H}_{s}^{\text{\emph{in}}}$ are rotationally
invariant, and as a consequence, their study is reduced to a study of their
parts $\hat{H}_{sj}^{\text{\emph{in}}}$ acting in $\mathfrak{H}_{sj}$, is
usually identified with the statement that the operators
$\hat{H}_{s}^{\text{\emph{in}}}$ commute with the respective generators
$\hat{J}_{s}$ of the rotation group, which in turn is often identified with
the commutativity of the differential operations $\check{H}_{s}$ and $\check{J}_{s}$,
$[\hat{H}_{s}^{\text{\emph{in}}},\hat{J}_{s}]=[\check{H}_{s},\check{J}_{s}]=0$.
Strictly speaking, such a statement is improper twice: the formal
commutativity of differential operations in no way implies the commutativity
of the associated operators, the more so as the commutator of two unbounded
operators in Hilbert space are generally not defined.

Each $\hat{H}_{sj}^{\text{\emph{in}}}$ is a symmetric operator in the
subspace $\mathfrak{H}_{sj}$. It evidently induces a symmetric operator
$\hat{h}_{\text{\emph{in}}}(Z,j,s)$ in the Hilbert space
$\mathbb{L}^{2}(\mathbb{R}_{+})$ that is unitarily equivalent to
$\hat{H}_{sj}^{\text{\emph{in}}}$,
$\hat{h}_{\text{\emph{in}}}(Z,j,s)\overset{def}{=}V_{sj}^{-1}\hat{H}_{sj}^{\text{\emph{in}}}V_{sj}$,
so that $\hat{h}_{\text{\emph{in}}}(Z,j,s)F=V_{s~j}^{-1}\hat{H}_{sj}^{\text{\emph{in}}}\Psi _{sj}$,
$\Psi_{sj}=V_{sj}F$. The $\hat{h}_{\text{\emph{in}}}(Z,j,s)$ is given by
\begin{equation}
\hat{h}_{\text{\emph{in}}}(Z,j,s)=\left\{
\begin{array}{l}
D_{h_{\text{\emph{in}}}(Z,j,s)}=\mathbb{C}_{0}^{\infty }(\mathbb{R}_{+}),\\
\hat{h}_{\text{\emph{in}}}(Z,j,s)F(\rho )=\check{h}(Z,j,s)F(\rho ),
\end{array}
\right.\label{3.13}
\end{equation}%
where $\mathbb{C}_{0}^{\infty }(\mathbb{R}_{+})=
C_{0}^{\infty }(\mathbb{R}_{+})\oplus C_{0}^{\infty }(\mathbb{R}_{+})$
and the differential operation $\check{h}(Z,j,s)$ reads
\begin{eqnarray}
&&\check{h}(Z,j,s)=-i\sigma_{y}\frac{d}{d\rho}+
\frac{\kappa }{\rho}\sigma_{x}-\frac{g}{\rho}+m\sigma_{z}, \nonumber \\
&&\kappa =-sj,\quad g=\alpha_{F}\epsilon ^{-1}Z, \label{3.14}
\end{eqnarray}
we call it the partial radial differential operation.

Constructing s.a. rotationally invariant Hamiltonians $\hat{H}_{s}$ as s.a.
extensions of the initial symmetric operators $\hat{H}_{s}^{\text{\emph{in}}}$, $\hat{H}_{s}=\hat{H}_{s}^{\mathfrak{e}}$, reduces to constructing s.a.
partial radial Hamiltonians $\hat{h}(Z,j,s)$ in $\mathbb{L}^{2}(\mathbb{R}_{+})$
as s.a. extensions of the initial symmetric partial radial operators
$\hat{h}_{\text{\emph{in}}}(Z,j,s)$, $\hat{h}(Z,j,s)=\hat{h}_{\mathfrak{e}}(Z,j,s)$.
This goes as follows. Let $\hat{h}_{\mathfrak{e}}(Z,j,s)$ be such
extensions, they evidently induce the s.a. extensions $\hat{H}_{sj}^{\mathfrak{e}}=V_{sj}\hat{h}_{\mathfrak{e}}(Z,j,s)V_{sj}^{-1}$
of the initial symmetric operators $\hat{H}_{sj}^{\text{\emph{in}}}$ in the subspaces
$\mathfrak{H}_{sj}$. Then the direct orthogonal sum of the partial operators
$\hat{H}_{sj}^{\mathfrak{e}}$,
\begin{equation}
\hat{H}_{s}^{\mathfrak{e}}=
\sum\limits_{j}{}^{\oplus}~\hat{H}_{sj}^{\mathfrak{e}},  \label{3.16}
\end{equation}%
is a s.a. rotationally invariant extension of the initial symmetric operator
$\hat{H}_{s}^{\text{\emph{in}}}$, the desired s.a. rotationally invariant
Hamiltonian $\hat{H}_{s}=\hat{H}_{s}^{\mathfrak{e}}$ in $\mathfrak{H}$.
Conversely, any s.a. rotationally invariant extension of the initial
symmetric operator $\hat{H}_{s}^{\text{\emph{in}}}$ has structure (\ref{3.16}).
The spectrum of the Hamiltonian $\hat{H}_{s}^{\mathfrak{e}}$ is a union
of the spectra of partial radial Hamiltonians,
\emph{spec~}$\hat{H}_{s}^{\mathfrak{e}}=\cup _{j}$\emph{spec~}$\hat{h}_{\mathfrak{e}}(Z,j,s)$,
and the corresponding eigenfunctions related to $\mathfrak{H}_{sj}$ are obtained
from the eigenfunctions of $\hat{h}_{\mathfrak{e}}(Z,j,s)$ in $\mathbb{L}^{2}(\mathbb{R}_{+})$
by the transformation $V_{sj}$, see (\ref{3.9}).

We already said above that in our consideration, we follow a similar
consideration for the s.a. rotationally invariant (with respect to the
\textrm{Spin}$(3)$ group) Dirac Hamiltonian for an electron in the Coulomb
field in three dimensions \cite{GVT, book2012, PhScr2013, VorGitLevFer2016}.
We recall that in \cite{GVT, book2012, PhScr2013, VorGitLevFer2016} there
was solved the problem of constructing and spectral analysis of s.a. partial
radial Hamiltonians $\hat{h}_{\mathfrak{e}}(Z,j,\zeta)$ in
$\mathbb{L}^{2}(\mathbb{R}_{+})$ as s.a. extensions of the initial partial radial
symmetric operators $\hat{h}_{\text{\emph{in}}}(Z,j,\zeta)$ defined on
$\mathbb{C}_{0}^{\infty }(\mathbb{R}_{+})$ and associated with the
differential operations
\begin{equation}
\check{h}(Z,j,\zeta)=-i\sigma _{y}\frac{d}{dr}+
\frac{\varkappa }{r}\sigma_{x}-\frac{q}{r}+m\sigma_{z},  \label{3.18}
\end{equation}
where $r=\sqrt{x^{2}+y^{2}+z^{2}}$, $\varkappa =\zeta j$, $j=1/2,3/2,\dots$
is the $3-$dimensional angular
momentum quantum number, $\zeta =\pm 1$ is the spin number, $q=\alpha Z$,
$\alpha$ is the fine structure constant, $m$ is the electron mass. The
differential operation (\ref{3.18}) differs from differential operation
$\check{h}(Z,j,s)$ given by Eq. (\ref{3.14}) only by values and an
interpretation of parameters involved. To make a comparison with the
three-dimensional problem, it is convenient to introduce the parameter
$\zeta =\zeta (j,s)=-s\,\mathrm{sgn}(j)=\pm 1$,
$\kappa =\zeta(j,s)\left\vert j\right\vert.$

\section{General solution of radial equations\label{S4}}

We now turn to the general solution of the system of two linear ordinary
differential equations
\begin{equation}
\check{h}(Z,j,s)F(\rho )=WF(\rho ),\quad F(\rho )=
\begin{pmatrix}
f(\rho ) \\
g(\rho )
\end{pmatrix},
\quad W\in \mathbb{C},  \label{4.1}
\end{equation}
which is necessary in future in evaluating spectra and eigenfunctions of
partial radial Hamiltonians; the system (\ref{4.1}) is sometimes called the
(stationary) partial radial Schr\"{o}dinger equation. Real values of $W$ are
denoted by $E$ in what follows. For our purposes, it is actually sufficient
to consider $W$ belonging to the upper complex half-plane, $W=E+iy$,
$y\geq 0$. What is more, the limit $W\rightarrow E+i0$, is of our main interest.

System (\ref{4.1}) in terms of $f(\rho)$ and $g(\rho)$ has the form:
\begin{eqnarray}
\frac{df}{d\rho }+\frac{\kappa }{\rho }f(\rho )-\left( W+m+\frac{g}{\rho }\right)
g(\rho ) &=&0,  \notag \\
\frac{dg}{d\rho }-\frac{\kappa }{\rho }g(\rho )+\left( W-m+\frac{g}{\rho }\right)
f(\rho ) &=&0.  \label{4.2`}
\end{eqnarray}
We call Eqs. (\ref{4.2`}) the radial equations. The radial equations for the
three-dimensional problem have the same form.

For completeness, we present the general solution of the radial equations
following the standard procedure \cite{Akh62,book2012}. We begin with a
change of variables,
\begin{eqnarray*}
f(\rho ) &=&z^{\Upsilon }e^{-z/2}\left[ Q(z)+P(z)\right] , \\
g(\rho ) &=&i\Lambda z^{\Upsilon }e^{-z/2}\left[ Q(z)-P(z)\right],\quad
z=-2iK\rho,
\end{eqnarray*}
where
\begin{eqnarray*}
&&\Upsilon ^{2}=\kappa ^{2}-g^{2},\quad W\pm m=r_{\pm }e^{i\phi _{\pm}},\quad
0\leq \phi _{\pm }\leq \pi ,\quad r_{\pm }\geq 0, \\
&&\Lambda =\sqrt{\frac{W-m}{W+m}}=\sqrt{\frac{r_{-}}{r_{+}}}e^{-i(\phi_{+}-\phi _{-})/2},
\quad K=\sqrt{W^{2}-m^{2}}=\sqrt{r_{+}r_{-}}e^{i(\phi_{+}+\phi _{-})/2}.
\end{eqnarray*}
In new variables, the system of radial equations (\ref{4.2`}) reads
\begin{eqnarray}
&&z\frac{d^{2}Q(z)}{dz^{2}}+(\beta -z)\frac{dQ(z)}{dz}-\alpha Q(z)=0,\quad 
P(z)=-\frac{1}{b_{+}}\left( z\frac{d}{dz}+\alpha \right) Q(z),  \notag \\ \notag
&&\beta =1+2\Upsilon ,\quad \alpha =\alpha _{+},\quad 
\alpha_{+}=\Upsilon + \frac{gW}{iK},\quad
b_{+}=\kappa +\frac{gm}{iK}.  \label{4.3}
\end{eqnarray}
The equation for the function $Q(z)$ is the well-known confluent
hypergeometric equation.

Let $\Upsilon \neq -n/2$, $n\in \mathbb{N}$. The general solution of the
confluent hypergeometric equation is then a linear combination of the
standard confluent hypergeometric functions $\Phi(\alpha ,\beta ;z)$ and
$\Psi (\alpha ,\beta ;z)$,
\begin{equation}
Q(z)=A\Phi (\alpha,\beta ;z)+ B\Psi (\alpha ,\beta ;z),\label{4.3a}
\end{equation}
where $A, B=\mathrm{const}$,
\begin{eqnarray*}
\Psi(\alpha ,\beta ;z)=\frac{\Gamma (1-\beta )}{\Gamma (\alpha -\beta +1)}
\Phi(\alpha ,\beta ;z)+\frac{\Gamma (\beta -1)}{\Gamma (\alpha )}z^{1-\beta}
\Phi(\alpha -\beta +1,2-\beta ;z).
\end{eqnarray*}
Then using the relations
\begin{eqnarray*}
&&\left( z\frac{d}{dz}+\alpha \right) \Phi (\alpha ,\beta ;z)=
\alpha \Phi(\alpha +1,\beta ;z), \\
&&\left( z\frac{d}{dz}+\alpha \right) \Psi (\alpha ,\beta ;z)=
\alpha (\alpha-\beta +1)\Psi (\alpha +1,\beta ;z), \\
&&\alpha -\beta +1=-\alpha _{-},\quad \alpha _{+}\alpha _{-}=b_{+}b_{-},
\quad a=\frac{\alpha _{+}}{b_{+}}, \\
&&\alpha _{-}=\Upsilon -\frac{gW}{iK},\quad b_{-}=\kappa -\frac{gm}{iK},
\end{eqnarray*}
we find that the general solution of system (\ref{4.3}) is given by
\begin{eqnarray*}
Q(z) &=&A\Phi (\alpha,\beta ;z)+B\Psi (\alpha ,\beta ;z), \\
P(z) &=&-A a\Phi(\alpha +1,\beta ;z)+B b_{-}\Psi (\alpha +1,\beta ;z).
\end{eqnarray*}
And finally, using the relations
\begin{eqnarray*}
\Phi (\alpha +1,\beta ;z)=e^{z}\Phi (\beta -\alpha -1,\beta ;-z),\quad
i\Lambda \frac{1+a}{1-a}=\frac{\kappa +\Upsilon }{g},
\end{eqnarray*}
we represent the general solution of radial equations (\ref{4.2`}) in the
following form:
\begin{eqnarray}
&&F=AX(\rho ,\Upsilon ,W)
+Bz^{\Upsilon }e^{-z/2}\left[ \Psi (\alpha ,\beta; z)\varrho _{+}-
b_{-}\Psi (\alpha +1,\beta; z)\varrho _{-}\right] ,  \notag \\
&&\varrho_{\pm }=\left( \pm 1, i\Lambda \right)^{T},  \label{4.4}
\end{eqnarray}
where the doublet $X(\rho, \Upsilon, W)$ is
\begin{eqnarray}
&&X(\rho, \Upsilon, W)=\frac{(-2iK/m)^{-\Upsilon}}{1-a}z^{\Upsilon}e^{-z/2} 
\left[ \Phi (\alpha ,\beta ;z)\varrho _{+} + a \Phi (\alpha +1,\beta;z)\varrho _{-}\right]  \notag\\
&&\qquad\quad\quad\quad\,=\frac{(m\rho )^{\Upsilon }}{2}\left[ \Phi _{+}(\rho ,\Upsilon ,W)+
\Phi_{-}(\rho ,\Upsilon ,W)~\Xi \right] d_{+},  \notag \\
&&\Phi _{+}(\rho ,\Upsilon ,W)=e^{iK\rho }\Phi (\alpha ,1+2\Upsilon ,-2iK\rho )+
e^{-iK\rho}\Phi (\alpha _{-},1+2\Upsilon ,2iK\rho ),  \notag \\
&&\Phi _{-}(\rho ,\Upsilon ,W)=\frac{1}{iK}\bigg{[} e^{iK\rho }
\Phi(\alpha ,1+2\Upsilon,-2iK\rho )-e^{-iK\rho }
\Phi (\alpha _{-},1+2\Upsilon ,2iK\rho )\bigg{]},
\notag \\
&&\ \Xi =\begin{pmatrix}
            0 && m+W \\
            m-W && 0
         \end{pmatrix},\
d_{\pm }=\left( 1,\frac{\kappa \pm \Upsilon }{g}\right)^{T}.  \label{4.5}
\end{eqnarray}
In what follows, we use some particular solutions of radial equations
(\ref{4.2`}) corresponding to a specific choice of the constants $A$ and $B$ and
the parameter $\Upsilon$. First, we introduce a new quantity$\Upsilon_{+}$ as follows:
\begin{gather}
\Upsilon_{+}=\Upsilon_{+}(g,j)=
\begin{cases}
\gamma =\sqrt{\kappa ^{2}-g^{2}}\geq 0, & g\leq |\kappa |, \\
i\sigma =i\sqrt{g^{2}-\kappa ^{2}},\quad \sigma >0, & g>|\kappa |.
\end{cases} \label{4.6}
\end{gather}
This quantity $\Upsilon_{+}$ as a function of $g$ has zeros at the points
$g=g_{\mathrm{c}}(j)=\left\vert \kappa \right\vert =\left\vert j\right\vert$.

In the case $\Upsilon_{+}\neq 0$ $(g\neq g_{\mathrm{c}}(j))$, we take two
linearly independent solutions $F_{1}$ and $F_{2}$ forming a
fundamental system of solutions of system (\ref{4.2`}),
\begin{eqnarray}
F_{1}(\rho ;W) &=&X(\rho ,\Upsilon _{+},W)=
(m\rho )^{\Upsilon_{+}}d_{+}+O(\rho ^{\Upsilon _{+}+1}), \quad \rho \rightarrow 0,\notag \\
F_{2}(\rho ;W) &=&X(\rho ,-\Upsilon _{+},W)\notag =
(m\rho )^{-\Upsilon_{+}}d_{-}+O(\rho ^{-\Upsilon _{+}+1}),\quad
\rho \rightarrow 0,\label{4.7}
\end{eqnarray}
it is remarkable that the both doublets $F_{1}$ and $F_{2}$ are real entire
in $W$. Their Wronskian is $\mathrm{Wr}(F_{1},F_{2})=-2\Upsilon _{+}g^{-1}$.
If \textrm{Im}$W>0$ and $\rho \rightarrow \infty$, the both doublets
$F_{1}(\rho ;W)$ and $F_{2}(\rho ;W)$ increase exponentially.

Another useful solution $F_{3}$ is given by (\ref{4.4}) with $A=0$,
$\Upsilon =\Upsilon_{+}$ and a special choice for $B=B(W)$,
\begin{eqnarray}
&&F_{3}(\rho ;W)=B(W)z^{\Upsilon }e^{-z/2}\bigg{[}
\Psi (\alpha ,\beta; z)\varrho _{+}-
b_{-}\Psi (\alpha +1,\beta ;z)\varrho _{-}\bigg{]}  \\ \notag
&&\qquad\qquad\,=\Gamma (-2\Upsilon _{+})F_{1}(\rho ;W)-
\frac{\omega (W)F_{2}(\rho ;W)}{\mathrm{Wr}(F_{1},F_{2})},
\end{eqnarray}
\begin{eqnarray}
&&\omega(W)=\frac{\Gamma (1+2\Upsilon _{+})\Gamma (-\alpha _{-})
\left[igK+(\kappa +\Upsilon _{+})(W+m)\right] }{g\Gamma (\alpha )\left[
igK+(\kappa -\Upsilon _{+})(W+m)\right] } 
\left( 2e^{-i\pi /2}\frac{K}{m}\right) ^{-2\Upsilon _{+}}=
-\mathrm{Wr}(F_{1},F_{3}), \\
&& B(W)=\frac{1}{2}\Gamma (-\alpha _{-})\left[ 1+\frac{(m+W)(\kappa
+\Upsilon _{+})}{igK}\right] \left( 2e^{-i\pi /2}\frac{K}{m}\right)^{-\Upsilon _{+}}.  \label{4.8`}
\end{eqnarray}%
If $\rm{Im}W>0$ and $\rho \rightarrow \infty$, the doublet $F_{3}(\rho; W)$
decreases exponentially (with a polynomial accuracy).

In the special case of $\Upsilon _{+}=\gamma =0$ $(g=g_{\mathrm{c}}(j))$
where the doublets $F_{1}$ and $F_{2}$ coincide, we will consider two
linearly independent solutions $F_{1}^{(0)}$ and $F_{2}^{(0)}$ and their
linear combination $F_{3}^{(0)}$,
\begin{eqnarray}
&&F_{1}^{(0)}(\rho ;W)=F_{1}(\rho ;W)|_{\gamma =0}=d_{+}|_{\gamma =0}+
O(\rho),\quad d_{+}|_{\gamma =0}=\left(1,\zeta (j,s)\right)^T,\, \rho \rightarrow 0,  \label{4.9} \\
&&F_{2}^{(0)}(\rho ;W)=\partial _{\gamma }F_{1}(\rho ;W)|_{\gamma =0}-
\frac{\zeta (j,s)}{g_{c}(j)}F_{1}^{(0)}(\rho ;W)=d_{0}(\rho )+O(\rho \log \rho
),\, \rho \rightarrow 0,  \label{4.10a} \\
&&d_{0}(\rho )=\left(
\log (m\rho )-\zeta (j,s)g_{c}^{-1}(j),\zeta (j,s)\log (m\rho )
\right)^T ,  \label{4.10b} \\
&&F_{3}^{(0)}(\rho ;W)=-\lim_{\gamma \rightarrow 0}
F_{3}(\rho;W)=F_{2}^{(0)}(\rho ;W)+f(W)F_{1}^{(0)}(\rho ;W),\,
F_{3}^{(0)}\in \mathbb{L}^{2}(\mathbb{R}_{+}),\label{4.11a} \\
&&f(W)=g_{c}(j)\omega^{(0)}(W)=\log (2e^{-i\pi /2}K/m)+\psi (-ig_{c}(j)WK^{-1})\notag \\
&&\qquad\quad+(\zeta(j,s)(W-m)+iK))(2g_{c}(j)W)^{-1}-2\psi (1),
\quad \psi(x)=\Gamma ^{\prime }(x)/\Gamma (x).  \label{4.11b} 
\end{eqnarray}
The corresponding Wronskians are
\begin{equation*}
\mathrm{Wr}(F_{1}^{(0)},F_{2}^{(0)})=g_{\mathrm{c}}^{-1}(j),\quad
\mathrm{Wr}(F_{2}^{(0)},F_{3}^{(0)})=-\omega ^{(0)},
\end{equation*}
where $\omega^{(0)}=\omega ^{(0)}(W)$.

\section{Self-adjoint radial Hamiltonians\label{S5}}

Here it is useful to recall what was said in the end of Sec. \ref{S3}.
Because radial differential operation $\check{h}(Z,j,s)$ (\ref{3.14})
coincides with radial differential operation $\check{h}(Z,j,\zeta )$ (\ref{3.18}),
arising in solving the $3$-dimensional Coulomb problem in \cite{book2012},
up to the replacement and reinterpretation of the parameters
$\kappa \rightarrow \varkappa $, $g\rightarrow q$, we can use some results in
Ref. \cite{GVT, book2012, PhScr2013, VorGitLevFer2016} concerning s.a.
partial radial Hamiltonians $\hat{h}_{\mathfrak{e}}(Z,j,\zeta )$ in
$\mathbb{L}^{2}(\mathbb{R}_{+})$ for defining and the spectral analysis of s.a.
partial radial Hamiltonians $\hat{h}_{\mathfrak{e}}(Z,j,s)$ in
$\mathbb{L}^{2}(\mathbb{R}_{+})$ under the replacement and an appropriate
reinterpretation of the parameters $\varkappa \rightarrow \kappa$,
$q\rightarrow g$.

Because all possible s.a. partial radial Hamiltonians
$\hat{h}_{\mathfrak{e}}(Z,j,s)$ are associated with the common
differential operation $\check{h}(Z,j,s)$ (\ref{3.14}), see just below, their definition
reduces to specifying their domains $D_{h(Z,j,s)}\subset$
$\mathbb{L}^{2}(\mathbb{R}_{+})$. Each operator
$\hat{h}_{\mathfrak{e}}(Z,j,s)$ is a s.a. extension of initial
symmetric operator $\hat{h}_{\mathrm{in}}( Z,j,s)$ (\ref{3.13})
associated with the differential operation $\check{h}(Z,j,s)$ (\ref{3.14}),
s.a. in the sense of Lagrange, and defined on the space
$\mathbb{C}_{0}^{\infty }(\mathbb{R}_{+})$
of smooth compactly supported doublets on the semiaxis $\mathbb{R}_{+}$.
Simultaneously, each $\hat{h}_{\mathfrak{e}}(Z,j,s)$ is a s.a.
restriction, maybe trivial, of the operator
$\hat{h}_{\mathrm{in}}^{+}(Z,j,s)$, the adjoint of
$\hat{h}_{\mathrm{in}}(Z,j,s)$, which is associated with the same
differential operation $\check{h}(Z,j,s)$,
(it is just the reason of that each $\hat{h}_{\mathfrak{e}}(Z,j,s)$ is associated
with one and the same differential operation $\check{h}(Z,j,s)$), and is defined
on the so-called natural domain $D_{\check{h}(Z,j,s)}^{\ast }(\mathbb{R}_{+})$ for
$\check{h}(Z,j,s)$ consisting of the doublets
$F(\rho)\in \mathbb{L}^{2}(\mathbb{R}_{+})$ absolutely continuous in
$\mathbb{R}_{+}$ and such that $\check{h}(Z,j,s) F(\rho)\in \mathbb{L}^{2}(\mathbb{R}_{+})$,
$D_{h_{\mathrm{in}}(Z,j,s)}\subset \ D_{h_{\mathfrak{e}}(Z,j,s)}\subseteq
D_{\check{h}(Z,j,s) }^{\ast }(\mathbb{R}_{+})$.
A definition of s.a. radial Hamiltonians $\hat{h}_{\mathfrak{e}}(Z,j,s)$ essentially
depends on the values of the parameters $Z$ and $j$, more specifically, on the
variable $\Upsilon_{+}$ (\ref{4.6}).

By definition, the variable $j$ is half-integer-valued, both positive and
negative, $j=\pm (n+1/2)$, $n\in \mathbb{Z}_{+}$, while the variable $Z$ is
nonnegative integer valued, $Z\in \mathbb{Z}_{+}$, so that we deal with
the lattice of physically meaningful values of these variables in the upper 
($j$, $Z$) half-plane. However, it seems convenient to consider the variable
$Z$ continuous lying on the nonnegative vertical semiaxis, $Z\in \mathbb{R}_{+}$,
and return to its natural integer values of necessity.
\begin{figure}[h]
\label{graph1}
\center{
\includegraphics[width=8.6cm]{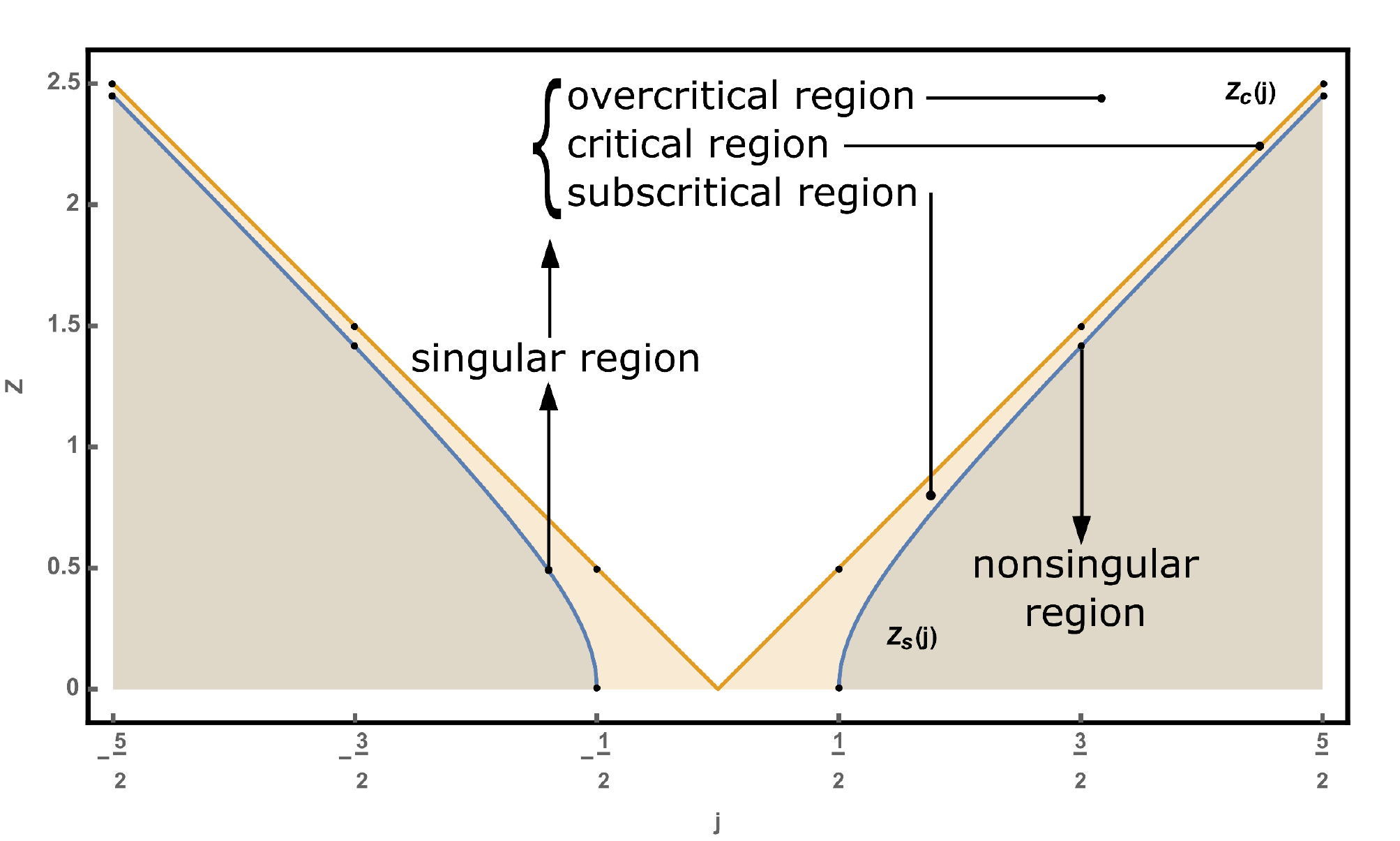}
}
\caption{Nonsingular and singular regions in the ($j$, $Z$) plane. The singular region consists of subcritical, critical, and overcritical subregions.
}
\end{figure}
There are the two regions in the upper ($j$, $Z$) half-plane (see Fig. 1 for $\alpha _{F}^{-1}\epsilon =1$), we call them the nonsingular
and singular ones, where the problem of s.a. extensions of
$\hat{h}_{\mathrm{in}}(Z,j,s)$ has principally different solutions: in the
nonsingular region, s.a. radial Hamiltonians $\hat{h}_{\mathfrak{e}}(Z,j,s)$
are defined uniquely, while in the singular region, they are defined nonuniquely.

These regions are separated by a symmetric singular curve $Z=Z_{\mathrm{s}}(j)$, where
\begin{equation*}
Z_{\mathrm{s}}\left( j\right) =\alpha _{F}^{-1}\epsilon ~\sqrt{j^{2}-
\frac{1}{4}},
\end{equation*}
which is equivalent to $g=g_{\mathrm{s}}(j)=\sqrt{j^{2}-1/4}$, or
$\Upsilon_{+}=\gamma = 1/2$, see (\ref{4.6}); this curve consists of two branches, the
right branch begins at the point ($j=1/2$, $Z=0$), and goes up toward the
right approaching a strait-line asymptote $Z=\alpha_{F}^{-1}\epsilon ~j$,
while the left branch begins at the point ($j=-1/2$, $Z=0$), and goes up
toward the left approaching a strait-line asymptote
$Z=-\alpha_{F}^{-1}\epsilon ~j$. The nonsingular region (the lower one) is defined by
the inequality $Z\leq Z_{\mathrm{s}}(j)$, which is equivalent
to the inequality $\Upsilon_{+}=\gamma \geq 1/2$, while the singular region
(the upper one) is defined by the inequality $Z>Z_{\mathrm{s}}(j)$,
which is equivalent to $0\leq \Upsilon_{+}=\gamma <1/2$ or
$\Upsilon_{+}=i\sigma$, $\sigma >0$, see (\ref{4.6}).

Note that the singular curve is an upper boundary of the nonsingular region.
The value $Z_{\mathrm{s}}(j)$ can be called the maximum nonsingular $Z$-value
for a given $j$. For illustration, we present the first $g_{\mathrm{s}}(j)$
values for first small half-integer $j$:
\begin{eqnarray*}
&&g_{\mathrm{s}}\left( \pm 1/2\right) =0,\quad
~g_{\mathrm{s}}\left( \pm 3/2\right) =\sqrt{2},\\
&&g_{\mathrm{s}}\left( \pm 5/2\right) =\sqrt{6},\quad
g_{\mathrm{s}}(\pm 7/2)=2\sqrt{3}.
\end{eqnarray*}
In what follows, we consider s.a. radial Hamiltonians $\hat{h}(Z,j,s)$
in the nonsingular and singular regions separately.

\subsection{Nonsingular region\label{S5.1}}

In the nonsingular region, $Z\leq Z_{\mathrm{s}}(j)$, each
partial radial Hamiltonian, we let $\hat{h}_{1}(Z,j,s)$ denote
it using subscript $1$ instead of $\mathfrak{e}$ as the symbol of the
nonsingular region (the subscripts $2$, $3$, $4$ together with other relevant
indices are used below instead of $\mathfrak{e}$ as symbols of particular
subregions of the singular region), is defined uniquely,
$\hat{h}_{1}(Z,j,s) =\hat{h}_{\mathrm{in}}^{+}( Z,j,s)$, in other
words, the initial symmetric operator $\hat{h}_{\mathrm{in}}(Z,j,s)$
is essentially s.a. because its deficiency indices are $(0,0)$,
and the domain of $\hat{h}_{1}(Z,j,s)$ is the natural domain
for $\check{h}(Z,j,s)$, $D_{h_{1}(Z,j,s)}=
D_{\check{h}(Z,j,s)}^{\ast}(\mathbb{R}_{+})$. Functions belonging to
$D_{\check{h}(Z,j,s)}^{\ast }(\mathbb{R}_{+})$ have the following asymptotic
behavior at the origin and at infinity:
\begin{equation*}
F(\rho )=O(\rho ^{1/2})\rightarrow 0,\;\rho \rightarrow 0;\
F(\rho)\rightarrow 0,\ \rho \rightarrow \infty.
\end{equation*}
Note that the points $j$, $Z=0$ with any $j$ belongs to the nonsingular
region, namely, to its lower boundary, which implies that the both total
free Dirac Hamiltonians $\hat{H}^{1}_{s}$, $s=\pm 1$, are defined uniquely as
s.a. operators in
$\mathfrak{H}=L^{2}(\mathbb{R}^{2})\oplus L^{2}(\mathbb{R}^{2})$.
Their spectra and (generalized) eigenfunctions are well known.

Turning to partial radial Hamiltonians $\hat{h}_{1}(Z,j,s)$ with $Z\neq 0$,
we first point out that\ the nonsingularity condition $\gamma ^{2}=j^{2}-g^{2}\geq 1/4$
excludes the values $j=\pm 1/2$ (we recall that the branches of the singular
curve begin at the points ($j=1/2$, $Z=0$) and ($j=-1/2$, $Z=0$)), and therefore, the
partial radial Hamiltonians $\hat{h}_{\mathfrak{e}}(Z,\pm 1/2,s)$ with $Z\neq 0$
are not uniquely defined.

Before proceeding to spectra and (generalized) eigenfunctions of uniquely
defined partial radial Hamiltonians $\hat{h}_{1}(Z,j,s)$ with
$Z\neq 0$, and therefore with $|j|>1/2$: $j=\pm 3/2$, $\pm 5/2,\dots$, we
recall some notions related to a classification of the spectrum points of a
s.a. operator following the conventional mathematical terminology, which
does not unfortunately coincide with the physical one.

We call eigenvalues of a s.a. operator only those points of its spectrum
which correspond to its bound states, the eigenstates whose wave functions
(doublets) are square integrable. Note that in the physical literature, any
spectrum point of a s.a. operator is often called its eigenvalue. Recall
that in the physical literature, eigenvalues of Hamiltonians are
conventionally called the energy levels.

The set of all eigenvalues of a given s.a. operator is called its point
spectrum. The set of all isolated eigenvalues of a given s.a. operator is
called its discrete spectrum, it is evidently a subset of the point spectrum
unless they coincide. In the physical literature, the point and discrete
spectrum are often identified.

The closure of the complement of the point spectrum in the whole spectrum of
a s.a. operator is called its continuous spectrum. In the physical
literature including textbooks, a rigorous definition of continuous spectrum
is usually absent, it is replaced by heuristic arguments and examples.

It can happen that a spectrum point of a s.a. operator belongs to its point
spectrum and to its continuous spectrum simultaneously. Such a spectrum
point is not a point of the discrete spectrum.

The spectrum of each partial radial Hamiltonian $\hat{h}_{1}(Z,j,s)$ in the
nonsingular region is simple (nondegenerate). It
consists of a continuous spectrum occupying the both negative and positive
semiaxis $(-\infty ,-m]$ and $[m,\infty)$ and of a discrete spectrum
located in the interval $(0,m)$ that consists of a growing infinite number
of energy levels $E_{n}$, accumulated at the point $m$,
\begin{eqnarray}
&&\mathrm{spec}\hat{h}_{1}(Z,j,s) = 
\left\{ E\in (-\infty ,-m]\cup
\lbrack m,\infty)\right\}\cup\left\{ E_{n}\in (0,m)\right\},\nonumber \\
&&E_{n}=E_{1n}(Z,j,s)=\frac{(n+\gamma )m}{\sqrt{g^{2}+(n+\gamma )^{2}}},\quad
n\in \mathcal{N}_{\zeta },\nonumber \\ \nonumber
&&   \\ \nonumber
&&\mathcal{N}_{\zeta }=
\begin{cases}
\mathbb{N}=\{1,2,\dots \}, & \zeta =1,\quad\text{\textrm{or }} sj < 0, \\
\mathbb{Z}_{+}=\{0,1,2,\dots \}, & \zeta =-1,\quad\text{\textrm{or }} sj > 0,
\end{cases}\\
&&\gamma =\sqrt{j^{2}-g^{2}}\geq 1/2,\quad
j = \pm 3/2,\pm 5/2,\dots. \label{5.1a}
\end{eqnarray}
The spectra of $\hat{h}_{1}(Z,j,s)$ can be obtained from the
spectra of the corresponding radial Hamiltonians $\hat{h}_{1}(Z,j,\zeta)$ in
the $3$-dimensional case (constructed in Refs.
\cite{GVT, book2012, PhScr2013, VorGitLevFer2016}) by the substitutions
\begin{eqnarray*}
&& j=1/2,3/2,\dots\rightarrow j=\pm 3/2,\pm 5/2,\dots,\\
&& \zeta =\pm 1\rightarrow \zeta =-s\,\mathrm{sgn}(j),\\
&&\gamma =\sqrt{(j+1/2)^{2}-q^{2}}\rightarrow \gamma =\sqrt{j^{2}-g^{2}}.
\end{eqnarray*}

We note that discrete energy levels with given $Z$ and~$j$ are independent
of $s$ and formally coincide for $s=\pm 1$, but the sets $\mathcal{N}_{\zeta}$,
the sequences of integers $n$ labelling the energy levels, are different
for different $s$, or equivalently, for different values of the variable
$\zeta =-s\,\mathrm{sgn}(j)=\pm 1$, namely, the sequences differ by first terms;
for brevity, we call these sets the sector $\zeta =+1$ and the sector $\zeta=-1$.

Normalized (generalized) eigenfunctions $U_{1E}(\rho)$,
$\left\vert E\right\vert \geq m$, of continuous spectrum and normalized
eigenfunctions $U_{1 n}(\rho)$ of bound states of energy $E_{n}$ for the partial
radial Hamiltonians $\hat{h}_{1}(Z,j,s)$ are given by
\begin{equation}
\begin{cases}
U_{1E}(\rho )=Q_{E}F_{1}(\rho ;E),~Q_{E}>0, & \left\vert E\right\vert \geq m,\\
U_{1n}(\rho )=Q_{n}F_{1}(\rho ;E_{n}),\quad n\in \mathcal{N}_{\zeta }, & 0<E<m,
\end{cases}  \label{5.2a}
\end{equation}
where the doublet $F_{1}(\rho ;E)$ is given by (\ref{4.5}) and (\ref{4.7}) and

\begin{eqnarray}
&&Q_{E}^{2}=\frac{2\pi g^{2}k\,\left( \left\vert E\right\vert -\mathrm{sgn}
(E)m\right) \left( 2k/m\right)^{2\gamma }e^{\pi gE/k}}{\Gamma^{2}(2\gamma +1)\left\vert
\Gamma \left( -\gamma +i\frac{g}{k}\left\vert E\right\vert \right) \right\vert^{2}\left(\cosh \left(2\pi \frac{g}{k}E\right)-\cos (2\pi \gamma ) \right)\left((\kappa +\gamma )^{2}k^{2}+g^{2}(E-m)^{2}\right)},\notag \\
&&Q_{n}^{2}=\frac{\Gamma \left( 2\gamma +1+n\right)\tau _{n}^{3}\left(
2\tau _{n}/m\right) ^{2\gamma }}{m^{2}n!\Gamma ^{2}(2\gamma +1)}\frac{ g\left( m-E_{n}\right) -
(\kappa-\gamma )\tau _{n} }{ g\left(
m-E_{n}\right) -(\kappa +\gamma )\tau_{n} },\notag \\ 
&&k=\sqrt{E^{2}-m^{2}},~\ \tau _{n}=gm[g^{2}+(n+\gamma )^{2}]^{-1/2},\label{5.2b}
\end{eqnarray}
they form a complete orthonormalized system in $\mathbb{L}^{2}(\mathbb{R}_{+})$
in the sense of inversion formulas, see \cite{book2012}.

In conclusion, we point out that a remarkable equality
$\hat{h}_{1}(Z,j,s)=\hat{h}_{1}(Z,-j,-s)$ takes place.

\subsection{Singular region\label{S5.2}}

In the singular region of the upper ($j$, $Z$) plane,
$Z>Z_{\mathrm{s}}(j)$, which is equivalent to $\Upsilon _{+}^{2}=j^{2}-g^{2}<1/4$, s.a.
partial radial Hamiltonians $\hat{h}_{\mathfrak{e}}(Z,j,s)$ as s.a.
extensions of the initial symmetric operators $\hat{h}_{\mathrm{in}}(Z,j,s)$
are not defined uniquely for each triple $Z$, $j$, $s$. The
reason is that the deficiency indices $m_{+}$, $m_{-}$ of each symmetric
operator $\hat{h}_{\mathrm{in}}(Z,j,s)$ are ($1$,$1$), and
therefore, there exists a one-parameter family of its s.a. extensions, s.a.
partial radial Hamiltonians, parametrized by the parameter
$\nu \in [-\pi /2,\pi /2]$, $-\pi /2\sim \pi /2$, its own for
each Hamiltonian. Partial radial Hamiltonians with the same triple $Z$, $j$,
$s$, but with different $\nu$ are associated with the same differential
operation $\check{h}(Z,j,s)$, but differ by their domains that are
subdomains of the natural domain $D_{\check{h}(Z,j,s)}^{\ast}
(\mathbb{R}_{+})$ for $\check{h}(Z,j,s)$ specified by certain asymptotic
s.a. boundary conditions at the origin which explicitly contain
the parameter $\nu$.

As in the $3$-dimensional Coulomb problem, we divide the singular region
into the three subregions, the respective subcritical, critical, and
overcritical regions, distinguished by a character of asymptotic s.a.
boundary conditions at the origin.

We let $\hat{h}_{k\nu }$, $k=2$, $3$, $4$, denote s.a. partial radial
Hamiltonians in the respective subcritical, $k=2$, critical, $k=3$, and
overcritical, $k=4$, regions (for brevity, we here omit their arguments $Z$,
$j$, $s$ of course, they are always implicitly implied). The s.a. boundary
conditions specifying Hamiltonians $\hat{h}_{k\nu }$ are similar in each
singular subregion, which provides a similar solution of the spectral
problem for $\hat{h}_{k\nu}$ with a given $k$.

For completeness, we briefly remind the reader of the basic points of the
spectral analysis of s.a. radial Hamiltonians $\hat{h}_{k\nu}$ based on the
Krein method of guiding functionals, for details, see \cite{book2012}.

We say in advance that in all three singular subregions, there exists a
unique guiding functional for each $\hat{h}_{k\nu}$, which implies that its
spectrum is simple (nondegenerate in physical terminology). In such a case,
the basic notion of spectral analysis is the spectral function $\sigma_{k\nu}(E)$,
$E\in\mathbb{R}$ is the energy variable, and especially its (generalized) derivative
$\sigma_{k\nu}^{\prime}(E)$. By construction, the function $\sigma_{k\nu}^{\prime}(E)$ is given by
\begin{equation*}
\sigma_{k\nu }^{\prime }(E)=\frac{1}{\pi }\mathrm{Im}\frac{1}{\omega_{k\nu}(E+i0)},
\end{equation*}
where the function $\omega_{k\nu}(W)$, $W=E+iy\in\mathbb{C}$, $y\neq 0$,
is the certain function which comes from the Green function of the operator
$\hat{h}_{k\nu}$, the kernel of the integral representation for the
resolvent $(\hat{h}_{k\nu}-W)^{-1}$ of the operator $\hat{h}_{k\nu}$,
namely, from a factor in the representation of the Green function in terms
of products of a doublet $U_{k\nu}(\rho; W)$ and the doublet $F_{3}(\rho; W)$
(\ref{4.8`}). The doublet $U_{k\nu}(\rho; W)$ is a linear
combination of doublets $F_{1}(\rho; W)$ and $F_{2}(\rho; W)$ (\ref{4.7})
that satisfies the asymptotic s.a. boundary conditions.

The spectrum of the operator $\hat{h}_{k\nu}$ is the support of the
function $\sigma_{k\nu}^{\prime}(E)$, and the restriction $U_{k\nu }(\rho; E)$
of the doublet $U_{k\nu}(\rho; W)$ to the spectrum point $E$ of the
operator $\hat{h}_{k\nu}$ is the corresponding eigenfunction of $\hat{h}_{k\nu}$.

In particular, a $\delta$ - function singularity of the function
$\sigma_{k\nu}^{\prime }$ at some point $E_{n}$ caused by a simple zero of the
real-valued function $\omega_{k\nu}(E)$ at this point, $\omega_{k\nu}(E_{n})=0$
and $\mathrm{Im}\omega_{k\nu}(E)=0$, $|E-E_{n}|<\delta$, is a
manifestation of the eigenvalue $E_{n}$ of the corresponding partial
Hamiltonian.

The points $E$ where the function $\omega_{k\nu}(E)$ is nonzero, not real,
and continuous are the points of the continuous spectrum of $\hat{h}_{k\nu}$.
At such points, the function $\sigma_{k\nu}^{\prime}(E)$ is positive,
$\sigma_{k\nu}^{\prime}(E)=Q_{k\nu}^{2}(E)>0$, and
$Q_{k\nu}(E)=\sqrt{\sigma_{k\nu}^{\prime}(E)}$ is the normalization factor for the
corresponding (generalized) eigenfunction $U_{k\nu}(\rho; E)$ of continuous
spectrum. We say in advance that in all three singular subregions, the
continuous spectrum of each $\hat{h}_{k\nu}$ is the union
$(-\infty,-m]\cup \lbrack m,\infty )$ of the two semiaxis.

In the interval $(-m,m)$, any function $\omega_{k\nu}(E)$ is real, but it
has isolated simple zeroes at some points $E_{kn}(\nu)$,
$\omega_{k\nu}(E_{kn}(\nu)) = 0$, $n=1,2,\dots$ (numbering can be different,
see below). These points are the isolated eigenvalues of the operator
$\hat{h}_{k\nu}$ forming its discrete spectrum, and the doublets
$U_{k\nu}(\rho; E_{kn}(\nu))$ are the corresponding (normalizable)
eigenfunctions. Really, in the vicinity of such points, the function
$1/\omega_{k\nu}(E+i0)$ is of the form
\begin{eqnarray*}
&&\frac{1}{\omega _{k\nu }(E+i0)}=-\frac{Q_{k\nu ,n}^{2}}{E-E_{kn}(\nu )+i0}
+O(1), \quad Q_{k\nu ,n}^{2}=-\frac{1}{\omega _{k\nu }^{\prime }\left( E_{kn}(\nu)\right) }>0,
\end{eqnarray*}
so that $\sigma_{k\nu}^{\prime}(E)=Q_{k\nu, n}^{2}\delta(E-E_{kn}(\nu))$,
and $Q_{k\nu, n} > 0$ is the normalization factor for the eigenfunction
$U_{k\nu}(\rho; E_{kn}(\nu))$. We say in advance that in all three singular
subregions, the discrete spectrum $\{E_{kn}(\nu)\}$ is a growing infinite
set of eigenvalues, energy levels, accumulated at the point $E=m$.

What is remarkable is that for each family $\{\hat{h}_{k\nu }(Z,j,s)\}$ of
partial radial Hamiltonians with a given $k$, there exists some value
$\nu=\nu_{-m}$ of the extension parameter such that
$\omega_{k\nu_{-m}}(-m)=0$, and the point $E=-m$ is an eigenvalue of
the operator $\hat{h}_{k\nu_{-m}}$ with the corresponding normalizable eigenfunction
$U_{k\nu_{-m}}(\rho; -m)$, i.e., belongs to the point spectrum of
$\hat{h}_{k\nu_{-m}}$. The peculiarity is that the point $E=-m$ also belongs to the
continuous spectrum of $\hat{h}_{k\nu_{-m}}$ being the upper boundary of its
lower branch. We thus encounter the case mentioned above where the
Hamiltonian $\hat{h}_{k\nu_{-m}}$ has a nontrivial point spectrum which is
not reduced to pure discrete one.

The normalized (generalized) eigenfunctions
$U_{k\nu, E}(\rho)=Q_{k\nu}(E)U_{k\nu}(\rho; E)$ of continuous spectrum
(in physical terminology, they \textquotedblleft are normalized to $\delta$ function
in energy\textquotedblright ) and normalized eigenfunctions
$U_{k\nu, n}(\rho)=Q_{k\nu, n}U_{k\nu}(\rho; E_{kn}(\nu))$ of discrete
spectrum form a complete orthonormalized system in
$\mathbb{L}^{2}(\mathbb{R}_{+})$ in the sense of inversion formulas.
We call the function $\omega_{\nu k}(E)$ the basic function and the
doublet $U_{k\nu}(\rho; W)$ the basic doublet.

Note that the spectral analysis in the nonsingular region (it was omitted
here, we present only its results) follow the same scheme with the basic
function $\omega (W)/\Gamma (-2\gamma )$, where the function $\omega (W)$ is
given in (\ref{4.8`}), instead of $\omega _{k\nu}(W)$ and basic doublet
$F_{1}(\rho; W)$ (\ref{4.7}) instead of $U_{k\nu}(\rho; W)$.

In what follows, we briefly outline the results of spectral analysis in each
singular subregion including a specification of partial radial Hamiltonians
$\hat{h}_{k\nu}$ and details of their point spectra.

\subsubsection{Subcritical region\label{S5.2.1}}

The subcritical region in the upper ($j$, $Z$) half-plane is defined by the
inequalities $Z_{\mathrm{s}}(j) <Z<Z_{\mathrm{c}}(j)$, which is equivalent to
$0<\Upsilon _{+}=\gamma <1/2$, where $Z_{\mathrm{c}}(j) =\alpha_{F}^{-1}\epsilon \left\vert j\right\vert$,
the latter is equivalent to $g=g_{\mathrm{c}}(j)=$ $\left\vert j\right\vert $,
or $\gamma = 0$, see Fig. 1.

The value $Z_{\mathrm{c}}(j)$ can be called the critical $Z$-value for a given
$j$. For illustration, we present the first $g_{\mathrm{c}}(j)$ values for first
small half-integer $j$:
\begin{equation*}
 g_{\mathrm{c}}\left( \pm 1/2\right) =0.5;
~g_{\mathrm{c}}\left( \pm 3/2\right) =1.5;
~g_{\mathrm{c}}\left( \pm 5/2\right) =2.5.
\end{equation*}

In the subcritical region, the s.a. radial Hamiltonians
$\hat{h}_{2\nu}(Z,j,s)$ are specified by asymptotic s.a. boundary conditions
at the origin of the form,
\begin{eqnarray}
&&F(\rho )=c[(m\rho )^{\gamma }d_{+}\cos \nu +(m\rho )^{-\gamma }d_{-}\sin \nu]+
O(\rho ^{1/2}),\nonumber \\ \label{4.1.1}
&&d_{\pm }=\left( 1,(\kappa \pm \gamma) /g\right)^{T},\quad
\rho \rightarrow 0,
\end{eqnarray}
where $c$ is an arbitrary complex number. The domain
$D_{h_{2\nu }(Z,j,s)}$ of the Hamiltonian $\hat{h}_{2\nu}(Z,j,s)$ is
\begin{eqnarray*}
D_{h_{2\nu }\left( Z,j,s\right) }=\big{\{} F(\rho ):\quad F(\rho )\in D_{\check{h}\left( Z,j,s\right) }^{\ast}
\left( \mathbb{R}_{+}\right) \text{ \textrm{and }}
F \,\mathrm{obey}\,(\ref{4.1.1})\big{\}}.
\end{eqnarray*}
The basic function $\omega_{2\nu}(W)$ and doublet $U_{2\nu}(\rho; W)$ are
the respective
\begin{equation*}
\omega_{2\nu}(W)=\frac{2\gamma}{g}\frac{\omega (W)\cos \nu + g^{-1}
\Gamma(1-2\gamma )\sin \nu }{\omega (W)\sin \nu -g^{-1}\Gamma (1-2\gamma )\cos \nu},
\end{equation*}
where the function $\omega (W)$ is given in (\ref{4.8`}), and
\begin{equation*}
U_{2\nu}(\rho; W) = F_{1}(\rho; W)\cos \nu +F_{2}(\rho; W)\sin \nu,
\end{equation*}
where the doublets $F_{1}(\rho; W)$ and $F_{2}(\rho; W)$ are given by (\ref{4.7}).
The derivative $\sigma_{2\nu}^{\prime}(E)$ of spectral function is given by
\begin{equation*}
\sigma _{2\nu }^{\prime }(E)=\frac{1}{\pi }\mathrm{Im}\frac{1}{\omega_{2\nu}(E+i0)}.
\end{equation*}
It is easy to determine the support of $\sigma_{2\nu}^{\prime}(E)$ and
find that the simple spectrum of the Hamiltonian $\hat{h}_{2\nu}(Z,j,s)$ is
given by
\begin{eqnarray*}
\mathrm{spec~}\hat{h}_{2\nu }\left( Z,j,s\right) =
\left\{E\in (-\infty,-m]\cup \lbrack m,\infty )\right\} \cup
\left\{ E_{2n}(\nu )\in \lbrack -m,m)\right\}.
\end{eqnarray*}
It consists of the continuous spectrum $(-\infty ,-m]\cup \lbrack m,\infty )$
and the point spectrum. The point spectrum is a growing infinite sequence
$\{E_{2n}(\nu)=E_{2n}(Z,j,s;\nu)\}$ of the energy levels $E_{2n}(Z,j,s;\nu)$
 that are the roots of the equation
\begin{eqnarray}
\omega_{2\nu }(E)  = \frac{2\gamma }{g}\frac{\omega (E)\cos \nu +g^{-1}
\Gamma(1-2\gamma )\sin \nu }{\omega (E)\sin \nu -g^{-1}\Gamma (1-2\gamma )\cos \nu}
=0,\quad E\in \lbrack -m,m), \label{4.1.2}
\end{eqnarray}
located in the semiinterval $[-m,m)$ and accumulated at the point $E=m$.
The infinite sequences $\{n\}$ of integers $n$ labelling the energy levels
depend on $s$ (compare with the nonsingular region) and are defined more
exactly below.

The normalized (generalized) eigenfunctions $U_{2\nu, E}(\rho)$ of
continuous spectrum and normalized eigenfunctions $U_{2\nu, n}(\rho)$ of
point spectrum given by
\begin{eqnarray}
&&U_{2\nu ,E}(\rho )=Q_{2\nu }(E)U_{2\nu }(\rho ;E)=Q_{2\nu }(E)\left(
F_{1}(\rho ;E)\cos \nu +F_{2}(\rho ;E)\sin \nu
\right), \notag \\
&&E\in (-\infty ,-m]\cup \lbrack m,\infty );  \notag \\
&&U_{2\nu ,n}(\rho )=Q_{2\nu ,n}U_{2\nu }\left( \rho ;E_{2n}(\nu )\right)=Q_{2\nu ,n}\big{(} F_{1}\left( \rho ;E_{2n}(\nu )\right) \cos \nu +
F_{2}\left( \rho ;E_{2n}(\nu )\right) \sin \nu \big{)}, \label{4.1.2b}
\end{eqnarray}
where
\begin{equation*}
Q_{2\nu }(E)=\sqrt{\sigma _{2\nu }^{\prime }(E)},\quad Q_{2\nu ,n}=
\sqrt{-\frac{1}{\omega _{2\nu }^{\prime }\left( E_{2n}(\nu )\right)}},
\end{equation*}
form a complete orthonormalized system in $\mathbb{L}^{2}(\mathbb{R}_{+})$
in the sense of inversion formulas.

Explicit expressions for the spectrum and eigenfunctions, including an
explicit solution of Eq. (\ref{4.1.2}), can be obtained in the two cases:
$\nu =\pm \pi /2$ and $\nu = 0$.

1) Let $\nu =\pm \pi /2$. In this case, we have
\begin{eqnarray*}
\left. \omega_{2\nu}(W)\right\vert _{\nu =\pm \pi /2} =
\frac{2\gamma }{g^{2}}\frac{\Gamma (1-2\gamma )}{\omega (W)},\quad
\left. U_{2\nu}(\rho, W)\right\vert_{\nu =\pm \pi /2}=F_{2}(\rho ;W).
\end{eqnarray*}
As we already mentioned above, the basic function and doublet for the
spectral analysis in the nonsingular region are
$\omega (W)/\Gamma (-2\gamma)$ and $F_{1}(\rho; W)$. But
\begin{equation*}
\frac{2\gamma }{g^{2}}\frac{\Gamma (1-2\gamma )}{\omega (W)}=
\left. \frac{\omega (W)}{\Gamma (-2\gamma )}\right\vert _{\gamma \rightarrow -\gamma },
\end{equation*}
as is easily verified, and
$F_{2}(\rho; W)=\left. F_{1}(\rho ;W)\right\vert_{_{\gamma \rightarrow -\gamma }}$.
It follows that all the results concerning the spectrum and eigenfunctions
in our case can be obtained from the corresponding results for the nonsingular
region, including (\ref{5.1a}), (\ref{5.2a}), by the formal substitution
$\gamma \rightarrow -\gamma$. In particular, the discrete spectrum is given by
\begin{equation}
\mathcal{E}_{2n}=E_{2n}(\pm \pi /2)=
\frac{(n-\gamma )m}{\sqrt{g^{2}+(n-\gamma )^{2}}},
\quad n\in \mathcal{N}_{\zeta },  \label{4.1.3}
\end{equation}
for $\mathcal{N}_{\zeta}$, see (\ref{5.1a}). A comment concerning an
implicit dependence of discrete energy levels $\mathcal{E}_{2n}$ on $s$, or
equivalently on $\zeta$, via the sequences $\mathcal{N}_{\zeta }$ of
integers $n$ labelling the energy levels is similar to that for formula (\ref{5.1a}).

2) Let $\nu =0$. In this case, we have
\begin{eqnarray*}
\left. \omega_{2\nu}(W)\right\vert _{\nu =0}=-\frac{2\gamma }
{\Gamma(1-2\gamma)}\omega(W) = \frac{\omega (W)}{\Gamma (-2\gamma )},\quad
\left. U_{2\nu}(\rho ,W)\right\vert _{\nu =0}=F_{1}(\rho ;W).
\end{eqnarray*}
It follows that in the case of $\nu =0$, all the results concerning the
spectrum and eigenfunctions are the direct extension of the corresponding
results, including (\ref{5.1a}), (\ref{5.2a}), (\ref{5.2b}) for the
nonsingular region where $\gamma \geq 1/2$ to the subcritical region where
$0<\gamma <1/2$.

A consideration of the general case $\left\vert \nu \right\vert <\pi /2$ is
completely similar to that for the singular region in the $3$-dimensional
Coulomb problem (see \cite{book2012}). We cite only most important
properties of the point spectrum. First, for each $Z$, $j$, $s$ and
$\nu \neq \nu_{-m}$, see (\ref{4.1.4}) just below, the point spectrum is a pure
discrete one. At $\nu =\nu_{-m}$, the discrete spectrum is complemented by
the energy level $E=-m$, which is simultaneously a point of continuous
spectrum, the upper boundary of its lower branch. This $\nu_{-m}$ is
determined from Eq. (\ref{4.1.2}) by setting $E=-m$ and noting that
$\omega(-m)=g^{-1}\Gamma(1+2\gamma )(2g)^{-2\gamma}$ to yield
\begin{equation}
\tan \nu _{-m}=-\frac{\Gamma (1+2\gamma )}{\Gamma (1-2\gamma )}
(2g)^{-2\gamma}.  \label{4.1.4}
\end{equation}%
It is remarkable that $\nu_{-m}=$ $\nu_{-m}(Z,j)$ depends on $|j|$ and is
independent of $s$.

For illustration, we give a graph (see Fig. 2) of the parameter $\nu_{-m}$ as function of $g$ for $j = \pm 1/2$.
\begin{figure}[h]
\label{nu01}
\center{
\includegraphics[width=4cm]{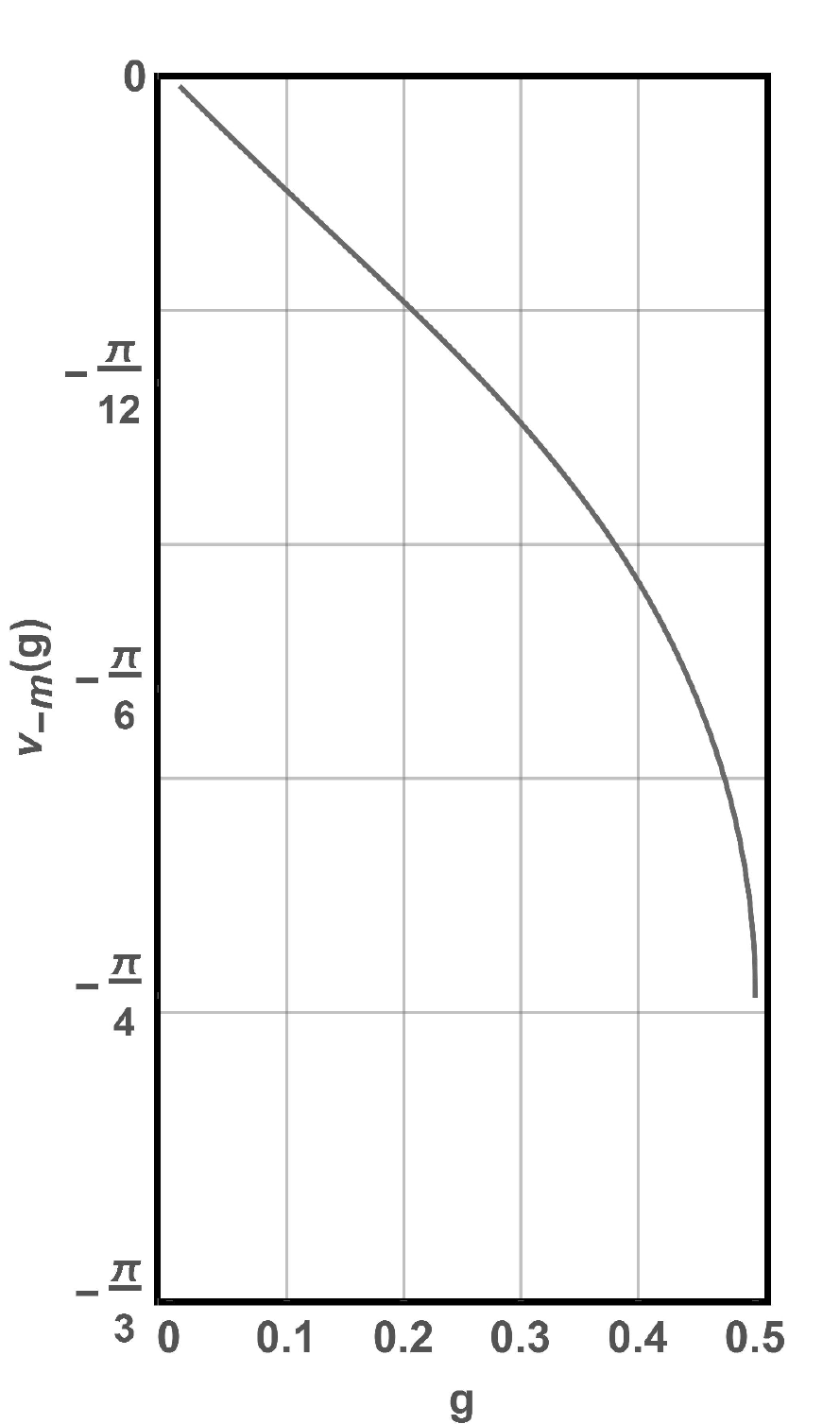}
}
\caption{$g$-dependence of $\protect\nu _{-m}$ for $j=1/2$.}
\end{figure}

Second, as follows from (\ref{3.14}) and is confirmed by (\ref{5.1a}) and
(\ref{4.1.3}), we should distinguish the sectors $\zeta =1$ and $\zeta =-1$
of energy levels, in particular, due to the different sequences of integers
$\mathcal{N}_{\zeta}$, see the comment to formula (\ref{5.1a}). What is
more, the dependence of the energy levels $E_{2n}(Z,j,s;\nu)$ on $s$ in the
general case of $0<|\nu |<\pi /2$ becomes explicit.

We let $n_{\zeta}$ denote the first term in the sequence
$\mathcal{N}_{\zeta}$: $n_{1}=1$, $n_{-1}=0$. It should be emphasized that in each
sector, there is the lowest energy level $-m$ corresponding to the same
$\nu_{-m}$ (\ref{4.1.4}), in particular $\mathcal{E}_{2n_{\zeta}}>-m$. We let
$E_{2(n_{\zeta}-1)}(\nu_{-m})$ denote this level.

In the energy semiinterval $[-m,\mathcal{E}_{2n_{\zeta }})$, there is one
energy level $E_{2(n_{\zeta }-1)}(\nu)$ for each $\nu \in (-\pi /2,\nu_{-m}]$
monotonically increasing from $-m$ to $\mathcal{E}_{2n_{\zeta }}-0$
as $\nu$ goes from $\nu_{-m}$ to $-\pi /2+0$, while for
$\nu \in (\nu_{-m},\pi /2)$, there is no energy level. In each energy interval
$\mathcal{E}_{2n}$, $\mathcal{E}_{2(n+1)}$, $n\geq n_{\zeta}$, for each
$\nu \in (-\pi/2,\pi /2)$ there is one energy level $E_{2n}(\nu )$ monotonically
increasing from $\mathcal{E}_{2n}+0$ to $\mathcal{E}_{2(n+1)}-0$ as $\nu$
goes from $\pi/2-0$ to $-\pi/2+0$. It is worth noting that
\begin{equation*}
\lim_{\nu \rightarrow -\pi/2}E_{2(n-1)}(\nu) =
\lim_{\nu \rightarrow \pi/2}E_{2n}(\nu) =
\mathcal{E}_{2n},\quad n\in \mathcal{N}_{\zeta}.
\end{equation*}

For illustration, we give graphs of low energy levels $j=1/2$ with $g=0.4$
as functions of $\nu $ for $s=+1$ (Fig. 3a) and $s=-1$ (Fig. 3b).
\begin{figure}[h]
\label{pic5.2.1.1}
\center{
\includegraphics[width=11cm]{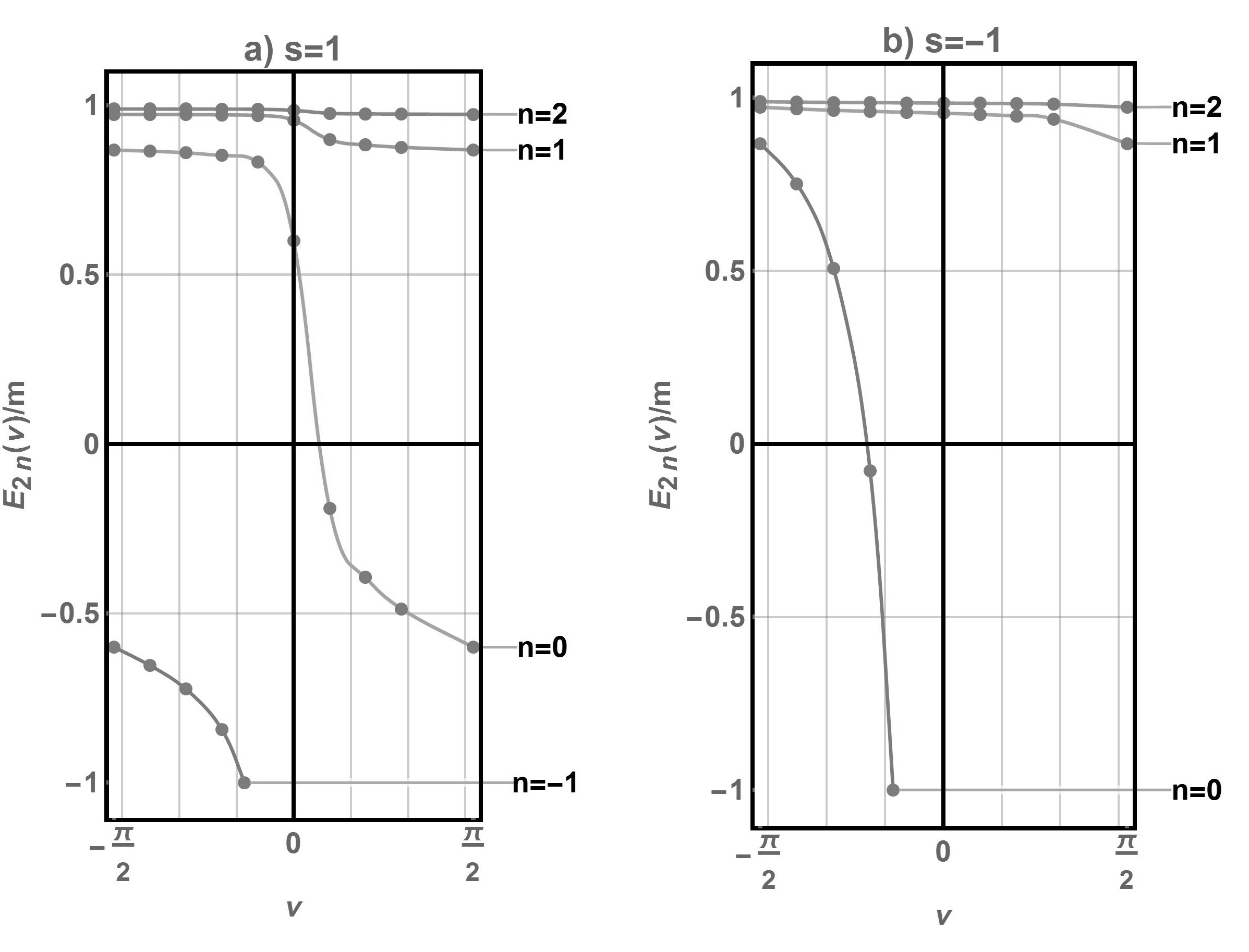}
}
\caption{$\nu$-dependence of energy levels $E_{2n}(\nu)$.}
\end{figure}

In conclusion, we point out that the remarkable equality
\begin{equation*}
\hat{h}_{2\nu }(Z,j,s)=\hat{h}_{2\nu }(Z,-j,-s)
\end{equation*}
holds, in particular, because s.a. boundary conditions (\ref{4.1.1}) are
invariant under the simultaneous replacements $j\rightarrow -j$, $s\rightarrow -s$.

\subsubsection{Critical region\label{S5.2.2}}

The critical region is the critical curve $Z=Z_{\mathrm{c}}(j)$, which is
equivalent to $g=g_{\mathrm{c}}(j)$, or $\Upsilon_{+} = \gamma = 0$. Note that
this region is of academic interest in some sense: the physical
values of the pairs $j$ (half-integer) and $Z$ (integer) lie on the critical
curve for very special values of the graphene \textquotedblright fine
structure constant \textquotedblright $\alpha_{F}/\epsilon$,
$\alpha_{F}/\epsilon =|j|/Z$. In particular, if $\alpha _{F}/\epsilon$ is an
irrational number, no physical pair ($j$, $Z$) lies on the critical curve. In
this region, the s.a. radial Hamiltonians $\hat{h}_{3\nu }(Z,j,s)$ are specified
by asymptotic s.a. boundary conditions at the origin of the form
\begin{eqnarray}
&&F(\rho )=c[d_{0}(\rho )\cos \nu +d_{+}\sin \nu ]+O(\rho ^{1/2}\log \rho
),\;\rho \rightarrow 0,  \notag \\
&&\nu \in \mathbb{[}-\pi /2,\pi /2],\quad -\pi /2\sim \pi /2,  \label{4.12a}
\end{eqnarray}
where the constant doublet $d_{+}=$ $d_{+}|_{\gamma =0}$ and the $\rho$ dependent
doublet $d_{0}(\rho)$ are given in respective (\ref{4.9}) and (\ref{4.10b}).
The domain $D_{h_{3\nu}(Z,j,s)}$ of the Hamiltonian $\hat{h}_{3\nu}(Z,j,s)$ is
\begin{eqnarray*}
D_{h_{3\nu }\left( Z,j,s\right) }=\big{\{} F(\rho ): \, F(\rho )\in D_{\check{h}\left( Z,j,s\right) }^{\ast}
\left( \mathbb{R}_{+}\right) \text{ \textrm{and }}
F\,\,\mathrm{obey}\,(\ref{4.12a})\big{\}}.
\end{eqnarray*}
The basic function $\omega_{3\nu}(W)$ is
\begin{equation*}
\omega _{3\nu }(W)=\frac{f(W)\cos \nu -\sin \nu }{g_{\mathrm{c}}(j)
\left[ f(W)\sin \nu +\cos \nu \right] },
\end{equation*}
where the function $f(W)$ is given by (\ref{4.11b}), and the basic doublet
$U_{3\nu}(W)$ is
\begin{equation*}
U_{3\nu }(W)=F_{1}^{(0)}(\rho ,W)\sin \nu +F_{2}^{(0)}(\rho ,W)\cos \nu,
\end{equation*}
where the doublets $F_{1}^{(0)}(\rho, W)$ and $F_{2}^{(0)}(\rho, W)$ are
given by the respective (\ref{4.9}) and (\ref{4.10a}), (\ref{4.10b}), while
the role of the doublet $F_{3}(\rho, W)$ plays the doublet $F_{3}^{(0)}(\rho, W)$
given by (\ref{4.11a}), (\ref{4.11b}). The derivative
$\sigma_{3\nu}^{\prime}(E)$ of the spectral function is given by
\begin{equation*}
\sigma _{3\nu }^{\prime }(E)=\frac{1}{\pi }\mathrm{Im}\frac{1}{\omega_{3\nu}(E+i0)}.
\end{equation*}

A knowledge of the function $\sigma_{3\nu}^{\prime}(E)$ requires a
knowledge of the reduction $f(E)=f(E+i0)$ of the function $f(W)$ (\ref{4.11a})
to the real axis. For $|E|\geq m$, it is given by
\begin{eqnarray}
f(E)=\ln \left( 2e^{-i\epsilon \pi /2}\frac{k}{m}\right) +\psi \left(
-ig_{\mathrm{c}}(j)\frac{|E|}{k}\right)
+\frac{\zeta (E-m)+i\epsilon k}{2g_{\mathrm{c}}(j)E}-2\psi (1),~|E|\geq m,\label{4.12b}
\end{eqnarray}
while for $|E|<m$, it is given by
\begin{eqnarray}
f(E)=\ln \left( 2\frac{\tau }{m}\right) +\psi \left( -g_{\mathrm{c}}(j)
\frac{E}{\tau }\right)+\frac{\zeta (E-m)-\tau }{2g_{\mathrm{c}}(j)E}-2\psi(1),\quad |E|<m, \label{4.12c}
\end{eqnarray}
where
\begin{eqnarray*}
k=\sqrt{E^{2}-m^{2}},\quad \tau =\sqrt{m^{2}-E^{2}},\quad
\epsilon =\mathrm{sgn}(E),\quad \zeta =\zeta (j,s),\quad
\psi(z)=\Gamma ^{\prime }(z)/\Gamma(z). 	
\end{eqnarray*}
Determining the support of $\sigma_{3\nu}^{\prime}(E)$ and evaluating the
relevant quantities (which is rather tedious) results in the conclusion that
the simple spectrum of the Hamiltonian $\hat{h}_{3\nu}(Z,j,s)$ is given by
\begin{eqnarray*}
\mathrm{spec~}\hat{h}_{3\nu}\left( Z,j,s\right) =
\left\{ E\in (-\infty,-m]\cup \lbrack m,\infty )\right\}\cup 
\left\{ E_{3n}(\nu )\in \lbrack -m,m)\right\},
\end{eqnarray*}
it consists of the continuous spectrum $(-\infty ,-m]\cup \lbrack m,\infty )$
and the point spectrum. The point spectrum is a growing infinite sequence
$\{E_{3n}(\nu )=E_{3n}(Z,j,s;\nu )\}$ of the energy levels $E_{3n}(Z,j,s;\nu)$
that are the roots of the equation $\omega_{3\nu }(E)=0$, $E\in\lbrack -m,m)$,
which is equivalent to
\begin{equation}
\frac{f(E)\cos \nu -\sin \nu }{f(E)\sin \nu +\cos \nu }=0,\quad |E|<m,
\label{4.13b}
\end{equation}
they are located in the semiinterval $[-m,m)$ and are accumulated at the
point $E=m$. The infinite sequences $\{n\}$ of integers $n$ labelling the
discrete energy levels depend on $s$ and are defined more exactly below.

It is also evident from (\ref{4.13b}) and (\ref{4.12c}) that energy levels
with given ($Z$, $j$) explicitly depend on $\zeta$, i.e., on $s$. The
normalized (generalized) eigenfunctions $U_{3\nu, E}(\rho)$ of continuous
spectrum and normalized eigenfunctions $U_{3\nu, n}(\rho)$ of point
spectrum given by
\begin{eqnarray}
&&U_{3\nu, E}(\rho) = Q_{3\nu}(E)U_{3\nu}(\rho; E)
= Q_{3\nu}(E)\left( F_{1}^{(0)}(\rho ;E)\cos \nu +F_{2}^{(0)}(\rho; E)
\sin \nu \right), \notag \\ \notag
&&E\in (-\infty ,-m]\cup \lbrack m,\infty );  \\ \notag
&&U_{3\nu, n}(\rho) = Q_{3\nu, n}U_{3\nu}\left(\rho; E_{3n}(\nu)\right)
=Q_{3\nu, n}\big{(} F_{1}^{(0)}\left( \rho; E_{2n}(\nu )\right) \cos \nu 
+F_{2}^{(0)}\left( \rho; E_{2n}(\nu)\right) \sin \nu \big{)}, \label{4.1.3c}
\end{eqnarray}
where
\begin{equation*}
Q_{3\nu }(E)=\sqrt{\sigma _{3\nu }^{\prime }(E)},\quad Q_{3\nu ,n}=
\sqrt{-\frac{1}{\omega _{3\nu }^{\prime }\left( E_{3n}(\nu )\right) }},
\end{equation*}
form a complete orthonormalized system in $\mathbb{L}^{2}(\mathbb{R}_{+})$
in the sense of inversion formulas.

In the case of $\nu =\pm \pi /2$, we obtain explicit formulas for the
spectrum and eigenfunctions. In this case, we have
\begin{eqnarray*}
&&\left. \omega _{3\nu}(W)\right\vert _{\nu =\pm \pi /2}=
-\frac{1}{g_{\mathrm{c}}(j)f(W)},\\
&&\left. U_{3\nu}(W)\right\vert_{\nu =\pm \pi /2} =
F_{1}^{(0)}(\rho ,W)=F_{1}(\rho ,W)|_{\gamma =0},\\
&&\left. \sigma_{3\nu}^{\prime }(E)\right\vert_{\nu =\pm \pi /2}=
-\frac{g_{\mathrm{c}}(j)}{\pi }\mathrm{Im} f(E+i0).
\end{eqnarray*}
In the range $|E|\geq m$ of the continuous spectrum, taking $f(E)$ (\ref{4.12b})
and using the relation $\psi (z)-\psi (-z)=-\pi \cot (\pi z)-z^{-1}$,
we have
\begin{eqnarray*}
\left. \sigma_{3\nu}^{\prime }(E)\right\vert_{\nu =\pm \pi /2}=
\left. Q_{3\nu}^{2}(E)\right\vert_{\nu = \pm \pi /2}
=\frac{g_{\mathrm{c}}(j)}{2}
\left( \coth (\pi g_{\mathrm{c}}(j)\frac{|E|}{k})+\epsilon \right),
\quad |E|\geq m.
\end{eqnarray*}
Note that
$\left. \sigma_{3\nu}^{\prime}(E)\right\vert_{\nu =\pm \pi/2}\geq 0$
for $|E|\geq m$, as it must, and the normalization factor
$\left. Q_{3\nu}^{2}(E)\right\vert_{\nu =\pm \pi /2}$ for the
eigenfunctions $\left. U_{3\nu}(\rho; E)\right\vert_{\nu =\pm \pi/2}=
F_{1}^{(0)}(\rho, E)$ of continuous spectrum is nonnegative.

In the range $|E|<m$, taking $f(E)$ (\ref{4.12c}) and having regard to that
the poles of $\left. \sigma _{3\nu}^{\prime}(E)\right\vert _{\nu =\pm \pi/2}$
determining the energy levels of the Hamiltonians are provided by the
poles of $\psi (-z)$ at the points $z_{n}=n$, $n\in \mathbb{Z}_{+}$, and the
pole of the third term with $\zeta =1$ in the r.h.s. of (\ref{4.12c}) at the
point $E=0$, we obtain that
\begin{equation*}
\left. \sigma _{3\nu }^{\prime }(E)\right\vert _{\nu =\pm \pi/2}=
\sum\limits_{n\in \mathcal{N}_{\zeta }}
\left. Q_{3\nu,n}^{2}\right\vert _{\nu =\pm \pi /2}\delta (E-\mathcal{E}_{3n}),
\end{equation*}
where the normalization factors
$\left. Q_{3\nu ,n}\right\vert _{\nu =\pm\pi /2}$ for the eigenfunctions
$\left. U_{3\nu }(\rho; \mathcal{E}_{3n})) \right\vert _{\nu =\pm \pi /2}=
F_{1}^{(0)}(\rho, \mathcal{E}_{3n})$ of bound states are
\begin{equation*}
\left. Q_{3\nu ,n}\right\vert _{\nu =\pm \pi /2}=\frac{\tau _{n}^{3/2}}{m},
~\tau_{n}=\frac{g_{\mathrm{c}}(j)m}{\sqrt{g_{\mathrm{c}}^{2}(j)+n^{2}}},
~n\in \mathcal{N}_{\zeta},
\end{equation*}
and the corresponding discrete energy levels
$\mathcal{E}_{3n}=E_{3n}(\nu=\pm \pi /2)$ are
\begin{equation*}
\mathcal{E}_{3n}=\frac{nm}{\sqrt{g_{\mathrm{c}}^{2}(j)+n^{2}}},
~n\in \mathcal{N}_{\zeta}.
\end{equation*}
Note that all the results for the spectrum and eigenfunctions in the
critical region, $\gamma = 0$, in the case of $\nu =\pm \pi /2$ are obtained
from the corresponding results in the nonsingular region, $\nu >1/2$,
including (\ref{5.1a}), (\ref{5.2a}), (\ref{5.2b}), by formally continuing
the latter to the point $\gamma = 0$.

As to the general case $|\nu |<\pi /2$, we cite only most important
properties of the discrete spectrum. An exposition is much like that in the
subcritical region.

First, for each $j$, $s$ in the critical region (where $Z$ is uniquely
determined by $j$), and $\nu \neq \nu_{-m}$, see (\ref{4.1.5}) just below,
the point spectrum is a pure discrete one. At $\nu =\nu_{-m}$, the discrete
spectrum is complemented by the energy level $E=-m$, which is simultaneously
a point of continuous spectrum, the upper boundary of its lower branch. This
$\nu _{-m}=$ $\nu _{-m}(|j|,\zeta )$ is determined from Eq. (\ref{4.13b}) by
setting $E=-m$ and noting that
$f(-m)=\ln (2g_{\mathrm{c}}(j))-2\psi(1)+\zeta /g_{\mathrm{c}}(j)$ to yield
\begin{equation}
\tan \nu _{-m}(|j|,\zeta )=\ln (2g_{\mathrm{c}}(j))-2\psi (1)+
\frac{\zeta }{g_{\mathrm{c}}(j)}.  \label{4.1.5}
\end{equation}%
It is remarkable that $\nu_{-m}$ as a function of $j$ and $\zeta$ depends
on $|j|$, while as a function of $j$ and $s$, it depends on the both $j$ and
$s$. It is also worth noting that $\nu_{-m}(|j|,\zeta)$ satisfies the
inequalities
\begin{eqnarray*}
&&\nu _{-m}(|j|,1)>\nu _{-m}(|j|,-1),\quad \nu _{-m}(|j|,1)>0,\quad \forall
j,  \notag \\
&&\nu _{-m}(1/2,-1)<0;\,\nu _{-m}(|j|,-1)>0,\quad |j|>3/2.
\end{eqnarray*}

Second, at given $j$, $\zeta$, in the energy semiinterval
$[-m,\mathcal{E}_{2n_{\zeta }})$ (we recall that $n_{\zeta}$ is equal to $1$
for $\zeta =1$ and to $0$ for $\zeta =-1$), for each
$\nu \in (-\pi /2,\nu_{-m}(|j|,\zeta)]$, $\nu_{-m}(|j|,\zeta )$ is given by
(\ref{4.1.5}), there is one energy level $E_{3(n_{\zeta }-1)}(\nu)$ monotonically
increasing from $-m$ to $\mathcal{E}_{3n_{\zeta}}-0$ as $\nu$ goes from
$\nu_{-m}(|j|,\zeta )$ to $-\pi/2+0$, while for
$\nu \in (\nu_{-m}(|j|,\zeta),\pi/2)$, there is no energy level. In each energy
interval $(\mathcal{E}_{3n}$, $\mathcal{E}_{3(n+1)})$, $n\geq n_{\zeta}$, for each
$\nu\in(-\pi/2,\pi/2)$, there is one energy level $E_{3n}(\nu)$ monotonically
increasing from $\mathcal{E}_{3n}+0$ to $\mathcal{E}_{3(n+1)}-0$ as $\nu$
goes from $\pi/2-0$ to $-\pi /2+0$. Note that relations
\begin{equation*}
\lim_{\nu \rightarrow -\pi /2}E_{3 (n-1)}(\nu)=
\lim_{\nu \rightarrow \pi/2}E_{3n}(\nu )=
\mathcal{E}_{3n},\quad n\in \mathcal{N}_{\zeta }
\end{equation*}
hold.

For illustration, we give graphs of low energy levels $j=1/2$ with $g=0.5$
as functions of $\nu $ for $s=+1$ (Fig. 4a) and for $s=-1$ (Fig. 4b).
\begin{figure}[h]
\label{fig3}
\center{
\includegraphics[width=9cm]{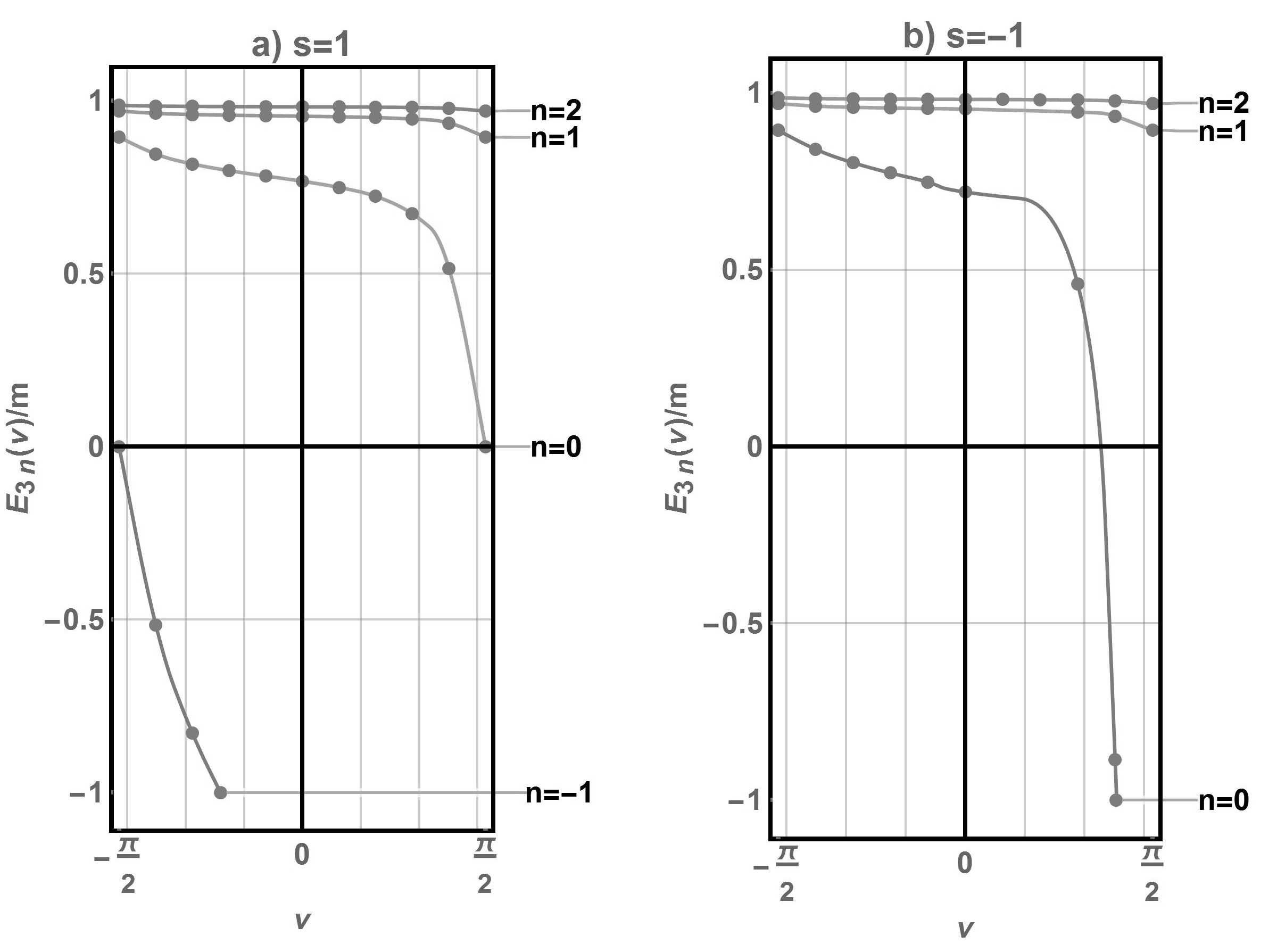}
}
\caption{$\nu$-dependence of energy levels $E_{3n}(\nu)$.}
\end{figure}

In conclusion, we point out that the remarkable equality
$\hat{h}_{3\nu}(Z,j,s)=\hat{h}_{3\nu }(Z,-j,-s)$ holds, in particular, because s.a.
boundary conditions (\ref{4.12a}) are invariant under the simultaneous
replacements $j\rightarrow -j$, $s\rightarrow -s$.

\subsubsection{Overcritical region\label{S5.2.3}}

The overcritical region in the upper ($j$, $Z$) half-plane is defined by the
inequality $Z>Z_{c}(j)=\alpha^{-1}\epsilon ~|j|$, which is equivalent to
$\Upsilon_{+}=i\sigma$, $\sigma =\sqrt{g^{2}-j^{2}}>0$. In this region, the
s.a. radial Hamiltonians $\hat{h}_{4\nu}(Z,j,s)$ are specified by the
asymptotic s.a. boundary conditions at the origin of the form
\begin{eqnarray}
&&F(\rho )=c\left( ie^{i\nu }(m\rho )^{i\sigma }d_{+}-ie^{-i\nu }
(m\rho)^{-i\sigma }d_{-}\right) + O(\rho^{1/2}), \quad
d_{\pm }=\left( 1,(\kappa \pm i\sigma)/g\right)^{T},\quad
\rho \rightarrow 0, \notag \\
&&\nu \in \lbrack -\pi /2,\pi /2],\quad -\pi /2\sim \pi /2. \label{5.2.3.0}
\end{eqnarray}
The domain $D_{h_{4\nu}(Z,j,s)}$ of the Hamiltonian $\hat{h}_{4\nu}(Z,j,s)$ is
\begin{eqnarray*}
D_{h_{4\nu}\left( Z,j,s\right) }=
\big{\{} F(\rho ): \quad F(\rho )\in D_{\check{h}\left( Z,j,s\right) }^{\ast}
\left( \mathbb{R}_{+}\right) \text{ \textrm{and }}
F\,\mathrm{obey}\,(\ref{5.2.3.0})\big{\}}.
\end{eqnarray*}
The basic function $\omega_{4\nu}(W)$ and doublet $U_{4\nu}(W)$ are given
by the respective
\begin{equation*}
\omega_{4\nu }(W)=-\frac{4i\sigma }{g}\frac{1-\frac{g}{\Gamma (1-2i\sigma )}
\omega (W)e^{2i\nu }}{1+\frac{g}{\Gamma (1-2i\sigma )}\omega (W)e^{2i\nu }},
\end{equation*}
where the function $\omega(W)$ is given in (\ref{4.8`}), and
$U_{4\nu}(W)=ie^{i\nu }F_{1}(\rho ;W)-ie^{-i\nu }F_{2}(\rho ;W)$, where the
doublets $F_{1}(\rho; W)$, $~F_{2}(\rho ;W)$ are given by (\ref{4.7}).
The derivative $\sigma_{4\nu }^{\prime}(E)$ of the spectral function is given by
\begin{equation*}
\sigma _{4\nu }^{\prime }(E)=\frac{1}{\pi}\mathrm{Im}
\frac{1}{\omega _{4\nu}(E+i0)}.
\end{equation*}

Determining the support of $\sigma_{4\nu }^{\prime }(E)$ and evaluating the
relevant quantities results in the conclusion that the simple spectrum of
the Hamiltonian $\hat{h}_{4\nu}(Z,j,s)$ is given by
\begin{eqnarray*}
\mathrm{spec~}\hat{h}_{4\nu}\left( Z,j,s\right)=
\left\{ E:\left\vert E\right\vert \geq m\right\} \cup 
\left\{ E_{4n}(\nu)\in \lbrack -m,m)\right\},
\end{eqnarray*}
it consists of the continuous spectrum $(-\infty ,-m]\cup \lbrack m,\infty )$
and the point spectrum. The point spectrum is a growing infinite sequence
$\{E_{4n}(\nu)=E_{4n}(Z,j,s;\nu)\}$ of the energy levels $E_{4n}(Z,j,s;\nu)$
that are the roots of the equation
\begin{eqnarray*}
\omega_{4\nu}(E)=-\frac{4i\sigma }{g}\frac{1-\frac{g}{\Gamma (1-2i\sigma )}
\omega (E)e^{2i\nu }}{1+\frac{g}{\Gamma (1-2i\sigma )}\omega (E)e^{2i\nu }}=0,\quad
E\in \lbrack -m,m),
\end{eqnarray*}
which is equivalent to the equation, we call it the spectral equation for
brevity,
\begin{equation}
1-\frac{g\omega (E)e^{2i\nu }}{\Gamma (1-2i\sigma )}=0,
\quad E\in \lbrack -m,m).
\label{5.2.3.1b}
\end{equation}
These energy levels are located in the semiinterval $[-m,m)$ and are
accumulated at the point $E=m$. The infinite sequences $\{n\}$ of integers
$n$ labelling the discrete energy levels are defined more exactly below. It
is evident that energy levels with given ($Z$, $j$) explicitly depend on $s$,
or $\zeta$, in view of an explicit dependence of the function $\omega (E)$
on $s$.

The spectral equation (\ref{5.2.3.1b}) allows another form more convenient
for a further analysis. It is sufficient to note that
\begin{eqnarray*}
-\frac{g\omega (E)}{\Gamma (1-2i\sigma )}=
\frac{\Gamma (2i\sigma )\Gamma(-i\sigma -g\frac{E}{\tau })
[\tau (\kappa +i\sigma )-g(m-E)]}{\Gamma(-2i\sigma )
\Gamma (i\sigma -g\frac{E}{\tau })[\tau (\kappa -i\sigma)-g(m-E)]}
\left( \frac{2\tau }{m}\right) ^{-2i\sigma }
=e^{-2i\Theta (E)},
\end{eqnarray*}
\begin{eqnarray*}
&&\Theta (E)=\sigma \ln \frac{2\tau }{m}+\frac{1}{2i}\bigg{\{}\ln \Gamma (-2i\sigma)-
\ln \Gamma (2i\sigma )+\ln \Gamma \left( i\sigma -g\frac{E}{\tau }\right)
-\ln \Gamma \left( -i\sigma -g\frac{E}{\tau }\right)\\
&&
\qquad\qquad\qquad\quad+\ln [\tau(g_{c}-i\zeta \sigma )-\zeta g(m-E)]-\ln [\tau (g_{c}+i\zeta \sigma )-
\zeta g(m-E)]\bigg{\}},
\end{eqnarray*}
and Eq. (\ref{5.2.3.1b}) becomes $\cos [\Theta (E)-\nu ]=0$. Note that
$\Theta(E)$ is a smooth function on $(-m,m)$.

The normalized (generalized) eigenfunctions $U_{4\nu ,E}(\rho )$ of
continuous spectrum and normalized eigenfunctions $U_{4\nu ,n}(\rho )$ of
discrete spectrum given by
\begin{eqnarray}
&& U_{4\nu, E}(\rho)=Q_{4\nu }(E)U_{4\nu }(\rho ;E) 
 = Q_{4\nu}(E)\left( ie^{i\nu }F_{1}(\rho ;E)-ie^{-i\nu }F_{2}(\rho; E)\right),\notag \\
&& E\in (-\infty ,-m]\cup \lbrack m,\infty );\notag\\
&&U_{4\nu, n}(\rho)=Q_{4\nu, n}U_{4\nu}\left(\rho; E_{4n}(\nu)\right)
 = Q_{4\nu, n}\bigg{(} i e^{i\nu }F_{1}\left(\rho; E_{4n}(\nu)\right)
-ie^{-i\nu}F_{2}\left( \rho; E_{4n}(\nu)\right) \bigg{)},  \label{5.2.3.1c}
\end{eqnarray}
where
\begin{equation*}
Q_{4\nu }(E)=\sqrt{\sigma _{4\nu }^{\prime }(E)},\quad Q_{4\nu ,n}=
\sqrt{-\frac{1}{\omega _{4\nu }^{\prime }\left( E_{4n}(\nu )\right) }},
\end{equation*}
form a complete orthonormalized system in $\mathbb{L}^{2}(\mathbb{R}_{+})$
in the sense of inversion formulas.

We now refine a description of the discrete spectrum.

First, for each $Z$, $j$, $s$ in the overcritical region and
$\nu \neq \nu_{-m}$, see (\ref{5.2.3.2b}) just below, the point spectrum is a
pure discrete one. At $\nu =\nu_{-m}$, the discrete spectrum is
complemented by the energy level $E=-m$, which is simultaneously a point of
continuous spectrum, the upper boundary of its lower branch. This $\nu _{-m}$
$(Z,j)$ is determined from Eq. (\ref{5.2.3.1b}) by setting $E=-m$ and noting
that $g\omega (-m)=\Gamma (1+2i\sigma )(2g)^{-2i\sigma}$ to yield
\begin{eqnarray}
e^{2i\nu_{-m}}=
e^{-i\pi }(2g)^{2i\sigma }\frac{\Gamma (-2i\sigma )}{\Gamma(2i\sigma )},
\quad \nu_{-m}\in \lbrack -\pi /2,\pi /2], \quad -\pi /2\sim \pi /2.\label{5.2.3.2b}
\end{eqnarray}
It is remarkable that $\nu_{-m}$ $(Z,j)$ does not depend on $s$.

For illustration, we give a graph (Fig. 5) of the parameter
$\nu_{-m}$ as function of $g$ for $j=1/2$.
\begin{figure}[h]
\label{nu02}
\center{
\includegraphics[width=4cm]{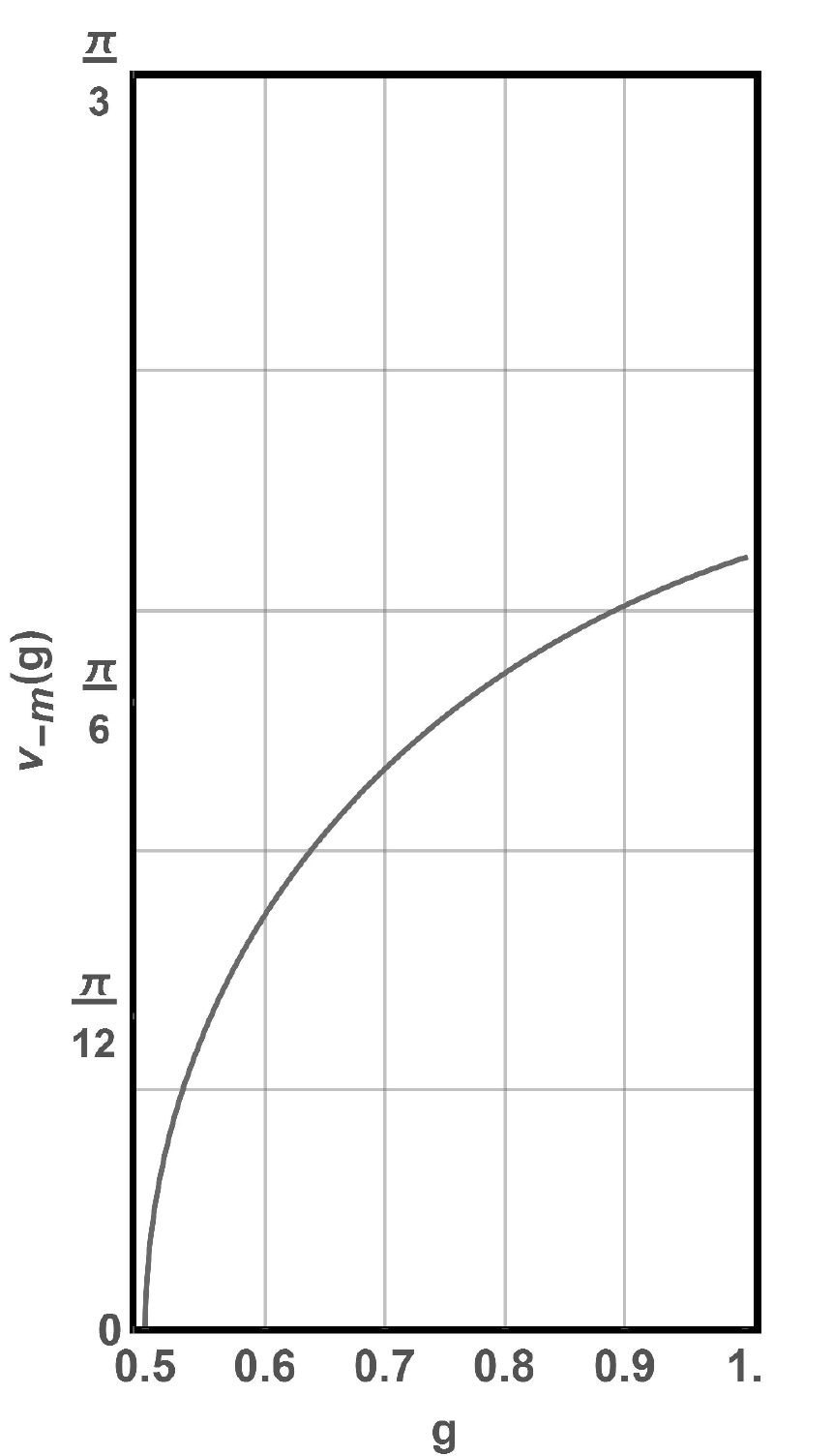}
}
\caption{$g$-dependence of $\nu_{-m}$ for $j=1/2$.}
\end{figure}

Second, in contrast to the previous regions, a labelling of the energy
levels in the overcritical region is independent of $\zeta $, i.e., it is
the same for $\zeta=1$ and $\zeta=-1$.

Third, to describe the point spectrum in this region, it appears convenient
to introduce a specific double labelling of the discrete energy levels with
$\nu =\pm \pi /2$. We formally distinguish values $\nu =-\pi/2$ and
$\nu=\pi/2$ labelling the energy levels with $\nu =-\pi /2$ by nonnegative
integers, $E_{4n}(-\pi /2)$, $n\in \mathbb{Z}_{+}=\{0,1,2,3,\dots\}$, while
the energy levels with $\nu =\pi /2$ are labelled by positive integers,
$E_{4n}(\pi /2)$, $n\in \mathbb{N}_{+}=\{1,2,3,\dots\}$. This artificial
difference is actually erased under the identification
$E_{4n}(-\pi/2)=E_{4(n+1)}(\pi/2)$, $n\in \mathbb{Z}_{+}$.

After this agreement, the point spectrum of Hamiltonians
$\hat{h}_{4\nu}(Z,j,s)$ looks as follows. In the energy interval
$[-m,E_{40}(-\pi /2)=E_{41}(\pi /2))$, there are no energy levels for
$\nu \in (\nu_{-m},\pi /2)$, while for any $\nu \in (-\pi /2,\nu _{-m}]$,
there is one energy level $E_{40}(\nu)$ which increases monotonically from
$-m$ to $E_{40}(-\pi /2)-0$ as $\nu$ goes from $\nu_{-m}$ to $-\pi /2+0$.
In each energy interval $[E_{4n}(\pi /2),E_{4(n+1)}(\pi/2)=E_{4n}(-\pi /2))$,
$n\in \mathbb{N}_{+}$, there is one energy level $E_{4n}(\nu)$, which
increases monotonically from $E_{4n}(\pi /2)$ to $E_{4n}(-\pi /2)-0$ as
$\nu$ goes from $\pi /2$ to $-\pi /2+0$. In particular, we have
$-m\leq E_{40}(-\pi /2) < E_{4n}(-\pi /2) < E_{4(n+1)}(-\pi /2) < m$,
$\forall n\in \mathbb{Z}_{+}$.

For illustration, we give graphs of low energy levels $j=1/2$ with $g=0.7$
as functions of $\nu $ for $s=+1$ (Fig. 6 a) and for $s=-1$ (Fig. 6 b).
\begin{figure}[h]
\label{fig4}
\center{
\includegraphics[width=9cm]{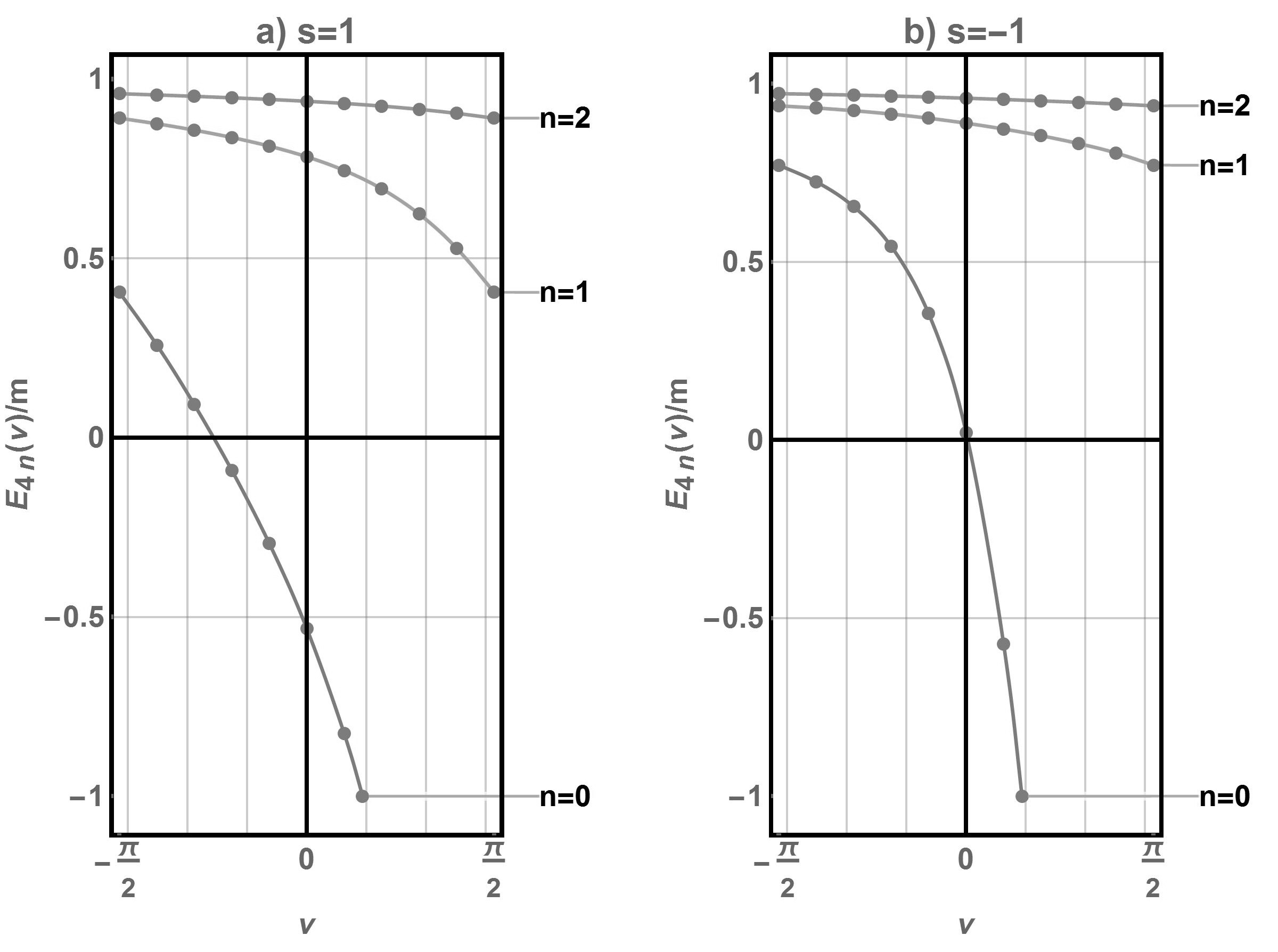}
}
\caption{$\nu$-dependence of energy levels $E_{4n}(\nu)$.}
\end{figure}

In conclusion, we point out that the remarkable equality
\begin{equation*}
\hat{h}_{4\nu }(Z,j,s)=\hat{h}_{4\nu }(Z,-j,-s)
\end{equation*}
holds, in particular, because s.a. boundary conditions (\ref{5.2.3.0}) are
invariant under the simultaneous replacements $j\rightarrow -j$, $s\rightarrow -s$.

\section{Self-adjoint total Hamiltonians and their spectra for point-like
case\label{S5.3}}

In sect. \ref{S5.1} and \ref{S5.2} we constructed all s.a. partial radial
Hamiltonians $\hat{h}_{\mathfrak{e}}(Z,j,s)$ for all values of $Z$ as s.a.
extensions of the initial symmetric operators $\hat{h}_{\text{\emph{in}}}(Z,j,s)$
for any $j$, $s$ and solved spectral problems for all these
Hamiltonians. As a result, (\ref{3.16}) allow one to restore all s.a.
operators $\hat{H}_{s}^{\mathfrak{e}}$ associated with the differential
operation (\ref{1.3b}) for any $g$ and to describe the solution of the
corresponding spectral problems for all the Hamiltonians $\hat{H}_{s}^{\mathfrak{e}}$.

It is convenient to introduce charge ranges in which the spectral problem
has a similar description. These ranges are defined by characteristic points
$k=l+1/2$, $l\in \mathbb{Z}_{+}$. The functions $g_{\mathrm{c}}(k)$ and
$g_{\mathrm{s}}(k)$ are defined on this set by
\begin{equation*}
g_{\mathrm{c}}(k)=k,\quad g_{\mathrm{s}}(k+1)=\sqrt{(k+1)^{2}-\frac{1}{4}},
\end{equation*}
and satisfy the following inequalities:
\begin{equation}
g_{\mathrm{c}}(k)<g_{\mathrm{s}}(k+1)<g_{\mathrm{c}}(k+1)<g_{\mathrm{s}}(k+2).
\label{5.3.1}
\end{equation}
Let us introduce intervals $\Delta (k)$ as follows:
\begin{eqnarray*}
&&\Delta (0)=\left( 0;\frac{1}{2}\right) ,\quad
\Delta (k)=(g_{\mathrm{c}}(k),g_{\mathrm{c}}(k+1)]=(k,k+1], \\
&&\left( 0,\infty \right) =\Delta (0)\cup \left\{ g_{\mathrm{c}}
\left( \pm \frac{1}{2}\right) \right\} \cup \left( \bigcup_{k}\Delta (k)\right).
\end{eqnarray*}
In turn, due to (\ref{5.3.1}), each interval $\Delta (k)$ can be represented
as $\Delta (k)=\cup _{i=1,2,3}\Delta _{i}(k)$, where
\begin{eqnarray*}
\Delta _{1}(k)=(g_{\mathrm{c}}(k),g_{\mathrm{s}}(k+1)],\quad
\Delta _{2}(k)=(g_{\mathrm{s}}(k+1),g_{\mathrm{c}}(k+1)),\quad
\Delta _{3}(k)=\{g_{\mathrm{c}}(k+1)\}.
\end{eqnarray*}
According to this division, we define three ranges
$G_{i}=\bigcup_{k}\Delta_{i}(k)$, $i=1,2,3$, of coupling parameters $g$,
such that any given $g>g_{\mathrm{c}}(\pm 1/2)=1/2$ generates a pair of
two integers, $k$ and $i=1,2,3$, such that $g\Longrightarrow (k,i)$, $g\in G_{i}$.
Then, as follows from the consideration represented in
Sect. \ref{S5.1} -- \ref{S5.2}, we obtain the following picture.

A. Let $g\Longrightarrow (k,1)$, that is, $g\in \Delta_{1}(k)$ for some $k$,
which means that
\begin{equation*}
k=g_{\mathrm{c}}(k)<g\leq g_{\mathrm{s}}(k+1)=\sqrt{(k+1)^{2}-\frac{1}{4}}.
\end{equation*}

Consider quantum numbers $\left\vert j\right\vert \leq k$. Then
$g>g_{\mathrm{c}}(k)\geq g_{\mathrm{c}}(j)$, which means that $g>g_{\mathrm{c}}(j)$.
Such quantum numbers $j$ are characteristic for the overcritical region
considered in Sect.\ref{S5.2.3}.

Consider quantum numbers $\left\vert j\right\vert \geq k+1$. Then
$g\leq g_{\mathrm{s}}(k+1)\leq g_{\mathrm{s}}(j)$. Such quantum numbers $j$ are
characteristic for the nonsingular region considered in Sect.\ref{S5.1}.

Therefore, for such coupling $g$, we have
\begin{eqnarray}
U_{E}(\rho )&=&
\begin{cases}
U_{4\nu ,E}(\rho ), & \left\vert j\right\vert \leq k, \\
U_{1E}(\rho ), & \left\vert j\right\vert \geq k+1,
\end{cases}
\quad \left\vert E\right\vert \geq m,  \notag \\
U_{n}(\rho )&=&
\begin{cases}
U_{4\nu ,n}(\rho ), & \left\vert j\right\vert \leq k, \\
U_{1n}(\rho ), & \left\vert j\right\vert \geq k+1,
\end{cases} \quad
E_{n}=
\begin{cases}
E_{4n}(\nu ), & \left\vert j\right\vert \leq k, \\
E_{1n}(Z,j,s), & \left\vert j\right\vert \geq k+1,%
\end{cases} \label{5.3.2}
\end{eqnarray}

B. Let $g\Longrightarrow (k,2)$, that is, $g\in \Delta _{2}(k)$ for some $k$,
which means that
\begin{equation*}
\sqrt{(k+1)^{2}-\frac{1}{4}}=g_{\mathrm{s}}(k+1)<g<g_{\mathrm{c}}(k+1)=k+1.
\end{equation*}

Consider quantum numbers $\left\vert j\right\vert \leq k$. Then
$g>g_{\mathrm{s}}(k+1)>g_{\mathrm{c}}(k)\geq g_{\mathrm{c}}(j)$, which means that
$g>g_{\mathrm{c}}(j)$. Such quantum numbers $j$ are characteristic for the
overcritical region considered in Sect.\ref{S5.2.3}.

Consider quantum numbers $\left\vert j\right\vert =k+1$. In this case
$g_{\mathrm{s}}(j)<g<g_{\mathrm{c}}(j)$. Such quantum numbers $j$ are
characteristic for the subcritical region considered in Sect.\ref{S5.2.1}.

Consider quantum numbers $\left\vert j\right\vert >k+1$. Then
$g<g_{\mathrm{c}}(k+1)<g_{\mathrm{s}}(k+2)\leq g_{\mathrm{s}}(j)$, so that
$g<g_{\mathrm{s}}(j)$. Such quantum numbers $j$ are characteristic for the nonsingular
region considered in Sect.\ref{S5.1}.

Therefore, we have
\begin{eqnarray}
U_{E}(\rho )&=&
\begin{cases}
U_{4\nu ,E}(\rho ), & \left\vert j\right\vert \leq k, \\
U_{2\nu ,E}(\rho ), & \left\vert j\right\vert =k+1, \\
U_{1E}(\rho ), & \left\vert j\right\vert >k+1,
\end{cases}%
\quad \left\vert E\right\vert \geq m,  \notag \\
U_{n}(\rho )&=&
\begin{cases}
U_{4\nu ,n}(\rho ), & \left\vert j\right\vert \leq k, \\
U_{2\nu ,n}(\rho ), & \left\vert j\right\vert =k+1, \\
U_{1n}(\rho ), & \left\vert j\right\vert >k+1,%
\end{cases}\quad
E_{n}=
\begin{cases}
E_{4n}(\nu ), & \left\vert j\right\vert \leq k, \\
E_{2n}(\nu ), & \left\vert j\right\vert =k+1, \\
E_{1n}(Z,j,s), & \left\vert j\right\vert >k+1.
\end{cases} \label{5.3.3}
\end{eqnarray}

C. Let $g\Longrightarrow (k,3)$, that is, $g\in \Delta_{3}(k)$ for some $k$,
which means that $g=g_{\mathrm{c}}(k+1)=k+1$.

Consider quantum numbers $\left\vert j\right\vert \leq k$. Then
$g=g_{\mathrm{c}}(k+1)>g_{\mathrm{c}}(k)\geq g_{\mathrm{c}}(j)$, so that
$g>g_{\mathrm{c}}(j)$. Such quantum numbers $j$ are characteristic for the
overcritical region considered in Sect. \ref{S5.2.3}.

Consider a quantum number $\left\vert j\right\vert =k+1$. Then
$g=g_{\mathrm{c}}(j)$. Such quantum numbers $j$ are characteristic for the critical region
considered in Sect. \ref{S5.2.2}.

Consider quantum numbers $\left\vert j\right\vert >k+1$.
Then $g=g_{\mathrm{c}}(k+1)<g_{\mathrm{s}}(k+2)\leq g_{\mathrm{s}}(j)$, so that
$g<g_{\mathrm{s}}(j)$. Such quantum numbers $j$ are characteristic for the nonsingular
region considered in Sect. \ref{S5.1}.

Therefore, for such charges, we have
\begin{eqnarray}
U_{E}(\rho )&=&
\begin{cases}
U_{4\nu ,E}(\rho ), & \left\vert j\right\vert \leq k, \\
U_{3\nu ,E}(\rho ), & \left\vert j\right\vert =k+1, \\
U_{1E}(\rho ), & \left\vert j\right\vert >k+1,
\end{cases}%
\quad \left\vert E\right\vert \geq m,  \notag \\
U_{n}(\rho )&=&
\begin{cases}
U_{4\nu ,n}(\rho ), & \left\vert j\right\vert \leq k, \\
U_{3\nu ,n}(\rho ), & \left\vert j\right\vert =k+1, \\
U_{1n}(\rho ), & \left\vert j\right\vert >k+1,
\end{cases} \quad
E_{n}=
\begin{cases}
E_{4n}(\nu ), & \left\vert j\right\vert \leq k, \\
E_{3n}(\nu ), & \left\vert j\right\vert =k+1, \\
E_{1n}(Z,j,s), & \left\vert j\right\vert >k+1.
\end{cases}\label{varfC}
\end{eqnarray}

D. Let $g\in \Delta (0)$, which means that $g<g_{\mathrm{c}}(\pm 1/2)$.
Consider quantum numbers $\left\vert j\right\vert >1/2$. Then
\begin{equation*}
g<g_{\mathrm{c}}\left( \pm \frac{1}{2}\right) =\frac{1}{2}<\sqrt{j^{2}-
\frac{1}{4}}=g_{\mathrm{s}}(j).
\end{equation*}
Such quantum numbers $j$ are characteristic for the nonsingular region
considered in Sect.\ref{S5.1}.

Consider quantum numbers $\left\vert j\right\vert =1/2$. Then
$0=g_{\mathrm{s}}(j)<g<g_{\mathrm{c}}(j)=1/2$. Such quantum numbers $j$ are characteristic
for the subcritical region considered in Sect.\ref{S5.2.1}.

Therefore, for such charges, we have
\begin{eqnarray}
U_{E}(\rho )&=&
\begin{cases}
U_{2\nu ,E}(\rho ), & \left\vert j\right\vert =1/2, \\
U_{1E}(\rho ), & \left\vert j\right\vert >1/2,
\end{cases}%
\quad \left\vert E\right\vert \geq m,  \notag \\
U_{n}(\rho )&=&
\begin{cases}
U_{2\nu ,n}(\rho ), & \left\vert j\right\vert =1/2, \\
U_{1\nu ,n}(\rho ), & \left\vert j\right\vert >1/2,
\end{cases}\quad
E_{n}=
\begin{cases}
E_{2n}(\nu ), & \left\vert j\right\vert =1/2, \\
E_{1n}(\nu ), & \left\vert j\right\vert >1/2.
\end{cases}\label{5.3.5}
\end{eqnarray}

E. Let $g=g_{\mathrm{c}}(\pm 1/2)$. Consider quantum numbers
$\left\vert j\right\vert >1/2$. Then
\begin{equation*}
g=g_{\mathrm{c}}\left( \pm \frac{1}{2}\right) =\frac{1}{2}<\sqrt{j^{2}-
\frac{1}{4}}=g_{\mathrm{s}}(j).
\end{equation*}
Such quantum numbers $j$ are characteristic for the nonsingular region
considered in Sect. \ref{S5.1}.

Consider quantum numbers $\left\vert j\right\vert =1/2$. Then
$g=g_{\mathrm{c}}(j)$. Such quantum numbers $j$ are characteristic for the critical region
considered in Sect. \ref{S5.2.2}.

Therefore, for such charges, we have
\begin{eqnarray}
U_{E}(\rho )&=&
\begin{cases}
U_{3\nu ,E}(\rho ), & \left\vert j\right\vert =1/2, \\
U_{1E}(\rho ), & \left\vert j\right\vert >1/2,
\end{cases}
\quad \left\vert E\right\vert \geq m,  \notag \\
U_{n}(\rho )&=&
\begin{cases}
U_{3\nu ,n}(\rho ), & \left\vert j\right\vert =1/2, \\
U_{1\nu ,n}(\rho ), & \left\vert j\right\vert >1/2,
\end{cases}\quad
E_{n}=
\begin{cases}
E_{3n}(\nu ), & \left\vert j\right\vert =1/2, \\
E_{1n}(\nu ), & \left\vert j\right\vert >1/2.
\end{cases}  \label{5.3.6}
\end{eqnarray}

We are now in a position to describe the spectral problem for all the s.a.
Dirac Hamiltonians with any coupling $g$. We note that the inequality
$g>g_{\mathrm{s}}(\pm 1/2)=0$ and (\ref{3.16}) implies an important fact: the
total s.a. Dirac Hamiltonian $\hat{H}_{s}$ is not uniquely determined for
any charge $Z=(\epsilon g)/\alpha _{F}$.

Consider eigenvectors $\Psi _{sj}(\mathbf{r})$ of any s.a. Dirac Hamiltonian
$\hat{H}_{s}^{\mathfrak{e}}$. They satisfy the following set of equations
(see Sect. \ref{S3}):
\begin{equation*}
\check{H}_{s}\Psi_{s j}(\mathbf{r})=E\,\Psi_{s j}(\mathbf{r}),\quad
\hat{J}_{s}\Psi_{s j}(\mathbf{r})=j\,\Psi_{s j}(\mathbf{r}),
\end{equation*}
where as the eigenvectors have the form
$\Psi _{sj}(\mathbf{r})=V_{sj}U_{E}(\rho)$, see (\ref{3.9}).

For any coupling constants $g$, the energy spectrum of any s.a. Dirac
Hamiltonian $\hat{H}_{s}^{\mathfrak{e}}$ contains the continuous spectrum
occupying both negative and positive semiaxis $(-\infty ,-m]$ and $[m,\infty)$
and of the discrete spectrum located in the interval $(0,m)$ and includes
a growing infinite number energy levels, accumulated at the point $E=m$.
Similar to the $3$-dimensional case (see. \cite{book2012}), the asymptotic
form as $n\rightarrow \infty$ of the spectrum is given by the well-known
nonrelativistic formula
\begin{equation*}
E_{n}^{\mathrm{nonrel}}=m-E_{n}=\frac{mg^{2}}{2n^{2}}.
\end{equation*}
The eigenfunctions $\Psi_{sjE}(\mathbf{r})$, $\left\vert E\right\vert \geq m$,
which correspond to the continuous part of the spectrum, are generalized eigenfunctions of $\hat{H}^{\mathfrak{e}}_{s}$, whereas the eigenfunctions $\Psi_{sjE_{n}}(\mathbf{r})$ of $\hat{H}_{s}^{\mathfrak{e}}$, which correspond to the of bound states of energy $E_{n}$, belong to the
Hilbert space $\mathfrak{H}$. Doublets $U_{E}(\rho)$ which correspond to
the of bound states of energy $E_{n}$, are denoted by $U_{E_{n}}(\rho)=U_{n}(\rho)$.
All the doublets $U_{E}(\rho)$ and $U_{n}(\rho)$ and the spectra $E_{n}$ depend on the extension
parameters, on the quantum numbers $j$, parameter $s$ and on the coupling $g$ according to (\ref{5.3.2})--(\ref{5.3.6}). It should be remembered that the extension parameters depend on
both $j$ and $s$.

\section{Spectral problem in regularized Coulomb field of the impurity\label{r}}

In this section, we consider the bound state and point spectrum problems
(for brevity, we speak about the point spectrum problem in what follows) for
an electron in graphene with an impurity generating an electric field with
a regularized Coulomb potential $V(\rho)$ given by
\begin{equation}
V(\rho )=-g\left\{
\begin{array}{c}
1/\rho_{0},\quad \rho \leq \rho _{0}, \\
1/\rho,\quad \rho \geq \rho _{0}.
\end{array}
\right.  \label{r4}
\end{equation}
Such a potential corresponds to the field of a positive charge impurity
distributed uniformly over the spherical surface of the cutoff radius $\rho_{0}$
(Fig. 7). The parameter $g$ is defined in Eq. (\ref{1.2b}).

We assume that the impurity is placed in the middle of the hexagon of the
graphene lattice. Regularization (\ref{r4}) represents a situation in which
the lattice parameter $a$ is the closest distance that electron hopping
between carbon sites can be from a potential source \cite{Pereira2}. Then,
we assume that the cutoff radius is of the order of the lattice constant $a$.
For numerical calculations, we use the value $\rho_{0}=0.6a$.
\begin{figure}[h]
\label{rpot}
\center{
\includegraphics[width=8cm]{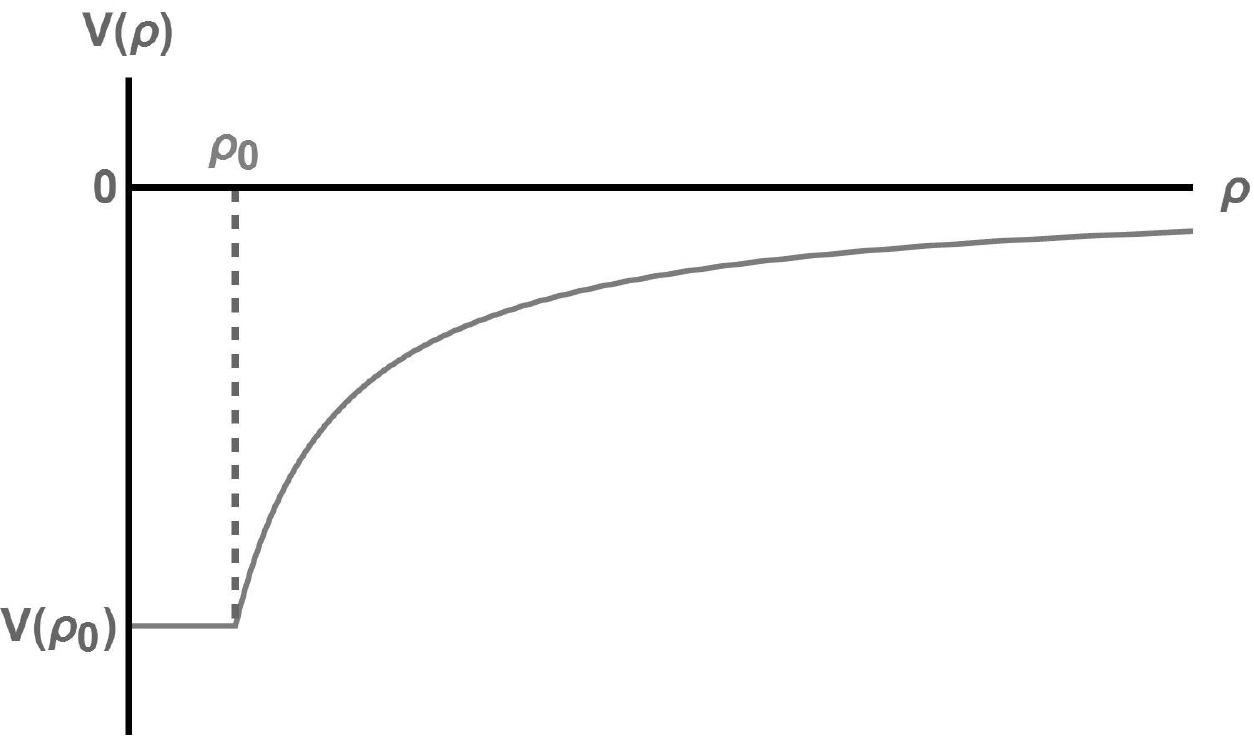}
}
\caption{Regularized Coulomb potential.}
\end{figure}
The Dirac Hamiltonian $\hat{H}_{s}^{\mathrm{reg}}$ is a differential operator in the Hilbert space
$\mathfrak{H=}\mathbb{L}^{2}(\mathbb{R}^{2})$, see Section \ref{S2}, associated with the
differential operation
\begin{equation*}
\check{H}_{s}^{\mathrm{reg}}=-i\left( s\sigma _{x}\partial _{x}+
\sigma_{y}\partial _{y}\right) +m\sigma _{z}+V(\rho)
\end{equation*}
and naturally defined. Each Hamiltonian $\hat{H}_{s}(Z)$ with any $Z$
is a uniquely defined s.a. operator as the sum of the uniquely defined free
Dirac Hamiltonian, see subsec. \ref{S5.1} and the bounded s.a. operator of
multiplication by a bounded real-valued function $V(\rho)$.

The Hamiltonian $\hat{H}_{s}^{\mathrm{reg}}$ is rotationally invariant from
the standpoint of the representation $U_{s}$ of the rotation group, and
therefore, similarly to the point-like case, see Section \ref{S3}, the point
spectrum problem for $\hat{H}_{s}^{\mathrm{reg}}$ is reduced to the point
spectrum problem for partial radial Hamiltonians $\hat{h}_{\mathrm{reg}}(Z,j,s)$
with fixed angular momenta $j=\pm 1/2,3/2,\dots$.
Each partial radial Hamiltonian $\hat{h}_{\mathrm{reg}}(Z,j,s)$ with fixed
$Z$, $j$, $s$ is an s.a. operator in the Hilbert space
$\mathbb{L}^{2}(\mathbb{R}_{+})=L^{2}(\mathbb{R}_{+})\oplus L^{2}(\mathbb{R}_{+})$ of doublets
$F(\rho)=( f(\rho), g(\rho) )^T$ ($f(\rho)$ and $g(\rho)$ are called the radial functions)
associated with the partial radial differential operation
\begin{equation*}
\check{h}_{\mathrm{reg}}(Z,j,s)=-i\sigma _{y}\frac{d}{d\rho }+
\frac{\kappa }{\rho }\sigma _{x}+V(\rho )+m\sigma _{z},
\end{equation*}
and defined on the natural domain for $\check{h}$: the domain $D_{h}$ of
each $\hat{h}$ consists of doublets $F(\rho )$ that are absolutely
continuous on $(0,\infty )$, vanish at the origin, $f(0)=g(0)=0$, and are
square integrable together with $\check{h}F(\rho )$ on $(0,\infty )$.

Because the potential $V(\rho)$ vanishes at infinity, the spectrum of each
partial radial Hamiltonian $\hat{h}_{\mathrm{reg}}(Z,j,s)$ consists of a
continuous part that is the union $(-\infty ,-m]\cup $ $[m,\infty )$ of two
half-axis and a point spectrum $\{E_{n}(Z,j,s), n\in\mathbb{Z}_{+}\}$
located in the segment $[-m,m]$. The total point spectrum
\emph{p.spec }$\hat{H}_{s}^{\mathrm{reg}}(Z)$ of the Dirac Hamiltonian
$\hat{H}_{s}^{\mathrm{reg}}(Z)$, which is the subject of our main interest,
is a union of the point spectra of partial radial hamiltonians
$\hat{h}_{\mathrm{reg}}(Z,j,s)$,
\begin{equation*}
\text{\emph{p.spec~}}\hat{H}_{s}^{\mathrm{reg}}(Z)=
\cup_{j}\text{\emph{p.spec~}}\hat{h}_{\mathrm{reg}}(Z,j,s),
\end{equation*}
while the corresponding eigenfunctions $\Psi _{sj}(\mathbf{r})$ of $\hat{H}_{s}(Z)$ in $\mathbb{L}^{2}(\mathbb{R}^{2})$ are obtained from the eigenfunctions $F(\rho )$ of $\hat{h}_{\mathrm{reg}}(Z,j,s)$ in $\mathbb{L}^{2}(\mathbb{R}_{+})$ by the unitary
transformation $V_{s~j}$, see (\ref{3.9}).

The point spectrum and the corresponding eigenfunctions of a partial radial
Hamiltonian $\hat{h}_{\mathrm{reg}}(Z,j,s)$ are defined by the solutions of
the stationary partial radial Schr\"{o}dinger equation
\begin{eqnarray*}
\hat{h}_{\mathrm{reg}}(Z,j,s)F(\rho )=EF(\rho ),\quad
E\in \lbrack-m,m],\quad F(\rho)\in D_{h(Z,j,s)}.
\end{eqnarray*}
This equation with fixed $Z$, $j$, and $s$ implies the system of differential
radial equations for the radial functions $f(\rho)$ and $g(\rho)$:
\begin{eqnarray}
&& f^{\prime }(\rho )+\frac{\kappa }{\rho }f(\rho )-k_{+}(\rho )g(\rho )=0, \notag \\
&& g^{\prime }(\rho )-\frac{\kappa }{\rho }g(\rho )+k_{-}(\rho )f(\rho )=0, \quad
 k_{\pm }(\rho )=E-V(\rho )\pm m.  \label{r1}
\end{eqnarray}
The system of radial equations (\ref{r1}) is supplemented by the condition
$-m\leq E\leq m$ and the following conditions on the functions $f(\rho)$ and
$g(\rho)$: they are absolutely continuous in $\rho$ on $(0,\infty)$,
satisfy the zero boundary conditions at the origin, $f(0)=g(0)=0$, and are
square integrable on $(0,\infty)$ (in fact, at infinity).

In finding the point spectra $\{E_{n}(Z,j,s)\}$ within the segment
$-m\leq E\leq m$, we must consider the open energy interval $-m<E<m$ and its
endpoints $E=m$ and $E=-m$ separately for technical reasons that become
clear below. We begin with the energy region $-m<E<m$.

\section{Discrete spectrum in $(-m,m)$\label{R1}}

\subsection{Solution of radial equations for $0\leq \protect\rho \leq $ $\protect\rho _{0}$\label{R1.1}}

In the interior region $0\leq \rho \leq \rho _{0}$, where we set
$f(\rho)=f_{\mathrm{in}}(\rho)$ and $g(\rho)=g_{\mathrm{in}}(\rho)$, the
functions $k_{\pm }(\rho )$ in Eqs. (\ref{r1}) and (\ref{r4}) are constants,
\begin{align*}
& k_{+}(\rho )=k_{+}=E+\frac{g}{\rho _{0}}+m,  \\
& k_{-}(\rho )=k_{-}=E+\frac{g}{\rho _{0}}-m,\quad E\in (-m,m).
\end{align*}
In this region, the first equation in (\ref{r1}) can be rewritten as
\begin{equation*}
g_{\mathrm{in}}(\rho )=\frac{1}{k_{+}}\left[ f_{in}^{\prime }(\rho )+
\frac{\kappa }{\rho }f_{in}(\rho )\right] .
\end{equation*}
Then, the second equation in (\ref{r1}) yields the second-order differential
equation for the function $f_{in}(\rho)$,
\begin{gather}
\frac{d^{2}f_{\mathrm{in}}(\rho )}{d\rho ^{2}}+\left( \eta ^{2}-
\frac{\nu^{2}-1/4}{\rho ^{2}}\right) f_{\mathrm{in}}(\rho )=0,\ \nu \in \mathbb{Z}_{+},  \notag \\
\eta =\sqrt{k_{+}k_{-}},\quad k_{+}k_{-}=
\left( E+\frac{g}{\rho _{0}}\right)^{2}-m^{2},  \quad
\nu =\left\vert \kappa +\frac{1}{2}\right\vert =|j|+\frac{\zeta }{2}=\left\{
\begin{array}{c}
|j|+1/2,~\zeta =1, \\
|j|-1/2,~\zeta =-1.
\end{array}
\right.   \label{r5b}
\end{gather}
Equation (\ref{r5b}) is supplemented by the two boundary conditions at the
origin:
\begin{equation*}
f_{\mathrm{in}}(0)=0,\ \left. [f_{\mathrm{in}}^{\prime }(\rho )+
\frac{\kappa}{\rho }f_{\mathrm{in}}(\rho)]\right\vert_{\rho = 0} = 0,
\end{equation*}
the latter is due to the condition $g_{\mathrm{in}}(0)=0$. Note that the
second condition is nontrivial only in the case $\nu =0$, if $\nu \neq 0$,
it follows from the first one.

The general solution of Eq. (\ref{r5b}) under the above boundary conditions
is (see \cite{Gradshteyn})
\begin{eqnarray}
f_{\mathrm{in}}(\rho )=c\sqrt{\rho }J_{\nu }(\eta \rho )=
c\sqrt{\rho}
\left\{
\begin{array}{l}
J_{|j|+1/2}(\eta \rho ),\ \mathrm{\ }\zeta =1, \\
J_{|j|-1/2}(\eta \rho ),\mathrm{\ }\zeta =-1,
\end{array}
\right. \quad c\in \mathbb{C}.  \label{r7a}
\end{eqnarray}
Using the relation
$J_{\nu }^{\prime }(z)\mp(\nu /z)J_{\nu }(z)=\mp J_{\nu \pm 1 }(z)$, see \cite{BatEr}, we find
\begin{eqnarray}
g_{\mathrm{in}}(\rho)=c\sqrt{\rho }\sqrt{\frac{k_{-}}{k_{+}}}
\zeta J_{\nu-\zeta }(\eta \rho )
=c\sqrt{\rho }\sqrt{\frac{k_{-}}{k_{+}}}\left\{
\begin{array}{l}
J_{|j|-1/2}(\eta \rho ),\ \mathrm{\ }\zeta =1, \\
-J_{|j|+1/2}(\eta \rho ),\mathrm{\ }\zeta =-1.
\end{array}
\right.   \label{r8a}
\end{eqnarray}

\subsection{Solution of radial equations for $\protect\rho _{0}\leq \protect\rho <\infty $\label{R1.2}}

In the exterior region $\rho \in \lbrack \rho _{0},\infty)$, where we set
$f(\rho )=f_{\mathrm{out}}(\rho)$, $g(\rho)=g_{\mathrm{out}}(\rho)$,
system (\ref{r1}), (\ref{r4}) is identical in form with the system of
equations in the $3$-dimensional Coulomb problem with a point charge, see
\cite{book2012} and \cite{VorGitLevFer2016}, the only difference is that
the charge parameter $g=\alpha _{F}Z/\epsilon$ is replaced by $\alpha Z$,
$\alpha$ is the fine structure constant, and the parameter $\kappa =-sj$,
$s=\pm 1$, $j=\pm 1/2,\pm 3/2,\dots$, is replaced by
$\varkappa=\zeta (j+1/2)$, $\zeta =\pm 1$, $j=1/2,3,2,\dots$.
The general solution of the latter system is well known.
In our case, system (\ref{r1}) and (\ref{r4}) is supplemented by the condition
that the functions $f_{\mathrm{out}}(\rho)$ and $g_{\mathrm{out}}(\rho)$ are
square integrable at infinity. Under this condition, a solution of the system is given by:
\begin{eqnarray}
&&f_{\mathrm{out}}(\rho )=B\sqrt{\frac{2m}{m-E}}(2\beta \rho )^{\mu}
e^{-\beta \rho }
\bigg{[}b_{-}\Psi (a+1,c;2\beta \rho )+\Psi (a,c;2\beta \rho )\bigg{]},
\notag \\
&&g_{\mathrm{out}}(\rho )=B\sqrt{\frac{2m}{m+E}}(2\beta \rho )^{\mu}
e^{-\beta \rho }
\bigg{[}b_{-}\Psi (a+1,c;2\beta \rho )-\Psi (a,c;2\beta \rho )\bigg{]},
\label{r9a}
\end{eqnarray}
where
\begin{eqnarray}
\beta =\sqrt{m^{2}-E^{2}},\quad \mu =\sqrt{\kappa ^{2}-g^{2}},\quad
a=\mu -\frac{gE}{\beta},\quad c=1+2\mu,\quad b_{-}=\kappa +\frac{gm}{\beta },
\label{r10}
\end{eqnarray}
and $\Psi$ is a standard confluent hypergeometric function vanishing at
infinity, see (\ref{4.3a}).

In what follows, we use Whittaker functions, see \cite{BatEr},
\begin{equation*}
W_{\lambda ,\mu }(x)=e^{-x/2}x^{c/2}\Psi (a,c;x),\ \lambda =\frac{c}{2}-a,
~\mu =\frac{c-1}{2},
\end{equation*}
and a new energy variable $\varepsilon $ defined as
$E=m\cos \varepsilon$, $\varepsilon = \arccos (E/m)\in (0,\pi )$.
In these terms, the final form of solutions of Eqs. (\ref{r1}) and (\ref{r4})
in the exterior region $\rho_{0}\leq \rho <\infty$ reads:
\begin{eqnarray*}
&& f_{\mathrm{out}}(\rho) = B\csc \left( \frac{\varepsilon }{2}\right)
(2\beta\rho )^{-1/2}
\bigg{[}(g\csc \varepsilon +\kappa )W_{\lambda ^{\prime },\mu }
(2\beta\rho )+W_{\lambda ,\mu }(2\beta \rho )\bigg{]}, \\
&& g_{\mathrm{out}}(\rho)=B\sec \left( \frac{\varepsilon }{2}\right)
(2\beta\rho )^{-1/2}
\bigg{[}(g\csc \varepsilon +\kappa )W_{\lambda ^{\prime },\mu }
(2\beta\rho )-W_{\lambda ,\mu }(2\beta \rho )\bigg{]},
\end{eqnarray*}
where
\begin{eqnarray*}
\beta = m\sin \varepsilon,\quad
\lambda = \frac{gE}{\beta }+\frac{1}{2}=g\cot\varepsilon + \frac{1}{2},\quad
\lambda^{\prime }=\lambda -1=g\cot \varepsilon - \frac{1}{2}.
\end{eqnarray*}

\subsection{Numerical solutions for discrete spectrum in the region $(-m,m)$, \label{R1.3}}

After the general solution of Eqs. (\ref{r1}) is found in the respective
regions $0\leq \rho \leq \rho _{0}$ and $\rho _{0}\leq \rho <\infty $, it
remains to satisfy the basic continuity condition for the solution as a
whole (to sew the partial solutions together smoothly), which reduces to the
requirement of continuity of the solution at the point $\rho =\rho _{0}$:
\begin{equation}
f_{\mathrm{in}}(\rho _{0})=f_{\mathrm{out}}(\rho _{0}),\quad
g_{\mathrm{in}}(\rho _{0})=g_{\mathrm{out}}(\rho _{0}).  \label{r16}
\end{equation}
If $c$ and $B$ are not zero, the compatibility of these conditions yields
the transcendental equation, which determines the point energy spectrum in
the region $(-m,m)$, which appears to be a discrete one in terms of the
variable $\varepsilon$,
\begin{eqnarray}
&&J_{\nu }(\eta \rho _{0})\sec \left( \frac{\varepsilon }{2}\right) \left[
(\kappa +g\csc \varepsilon )W_{\lambda ^{\prime },\mu }(2\beta \rho_{0})-
W_{\lambda ,\mu }(2\beta \rho _{0})\right]  \notag \\
&& -\sqrt{\frac{k_{-}}{k_{+}}}\zeta J_{\nu -\zeta }(\eta \rho _{0})
\csc\left( \frac{\varepsilon }{2}\right) \left[ \left( \kappa +g\csc \varepsilon\right)
W_{\lambda ^{\prime },\mu }(2\beta \rho _{0})+W_{\lambda ,\mu}(2\beta \rho _{0})\right]
= 0,  \label{r17}
\end{eqnarray}
with $k_{\pm }=m(\cos \varepsilon \pm 1)+g/\rho_{0}$. We call this basic
equation the spectrum equation for the interval $(-m,m)$. Strictly speaking,
we deal with a series of exact spectrum equations for given $Z$, $j$ and $s$.

After the spectrum equation is solved, i.e., the energy eigenvalues
$E_{n}(Z,j,s)$ of the partial radial Hamiltonians
$\hat{h}_{\mathrm{reg}}(Z,j,s)$ are found, the corresponding eigenfunctions
$F_{n}(Z,j,s)\in $ $\mathbb{L}^{2}(\mathbb{R}_{+})$ of bound states of Hamiltonians
$\hat{h}_{\mathrm{reg}}(Z,j,s)$ are obtained by the substitution of the evaluated
energy eigenvalues $E_{n}(Z,j,s))$ for $E$ in formulas (\ref{r7a}), (\ref{r8a}),
and (\ref{r9a}) with due regard to continuity condition (\ref{r16}),
according to which a unique normalization constant for wave functions
(doublets) remains undetermined. An analytical solution of the spectrum
equation (\ref{r17}) with any $Z$, $j$, $s$ is beyond the scope of our
possibilities. Only numerical solution of these equations is realizable at
present.

An equivalent expanded form of the spectrum equation (\ref{r17}) with the
certain fixed $\zeta$, which is maybe more suitable for numerical
calculations, is
\begin{eqnarray*}
\sqrt{\frac{k_{-}}{k_{+}}}\frac{J_{\left\vert j\right\vert -1/2}
(\eta \rho_{0})}{J_{\left\vert j\right\vert +1/2}(\eta \rho _{0})}
-\tan \left( \frac{\varepsilon }{2}\right)
\frac{\left[ \left( \left\vert j\right\vert +g\csc\varepsilon \right)
W_{\lambda ^{\prime },\mu }(2\beta \rho _{0})-W_{\lambda,\mu }(2\beta \rho _{0})
\right] }{
\left[ \left( \left\vert j\right\vert + g\csc \varepsilon \right)
W_{\lambda ^{\prime },\mu }(2\beta \rho_{0})+
W_{\lambda ,\mu }(2\beta \rho _{0})\right] }
=0,
\end{eqnarray*}
for $\zeta = 1$ and
\begin{eqnarray*}
\sqrt{\frac{k_{-}}{k_{+}}}\frac{J_{\left\vert j\right\vert +1/2}
(\eta \rho_{0})}{J_{\left\vert j\right\vert -1/2}(\eta \rho _{0})}
+\tan \left( \frac{\varepsilon }{2}\right) \frac{
\left[ \left( -\left\vert j\right\vert +g\csc\varepsilon \right)
W_{\lambda ^{\prime },\mu }(2\beta \rho _{0})-W_{\lambda,\mu }
(2\beta \rho _{0})\right] }{\left[ \left( -\left\vert j\right\vert+
g\csc \varepsilon \right) W_{\lambda ^{\prime },\mu }(2\beta \rho_{0})+
W_{\lambda ,\mu }(2\beta \rho _{0})\right] }=0,
\end{eqnarray*}
for $\zeta = -1$.

Numerical calculations for special case $j=1/2$, $s=1$, which produces the
lowest energy levels, are shown in Fig. 8.
\begin{figure}[h]
\label{reg_spec}
\center{
\includegraphics[width=9cm]{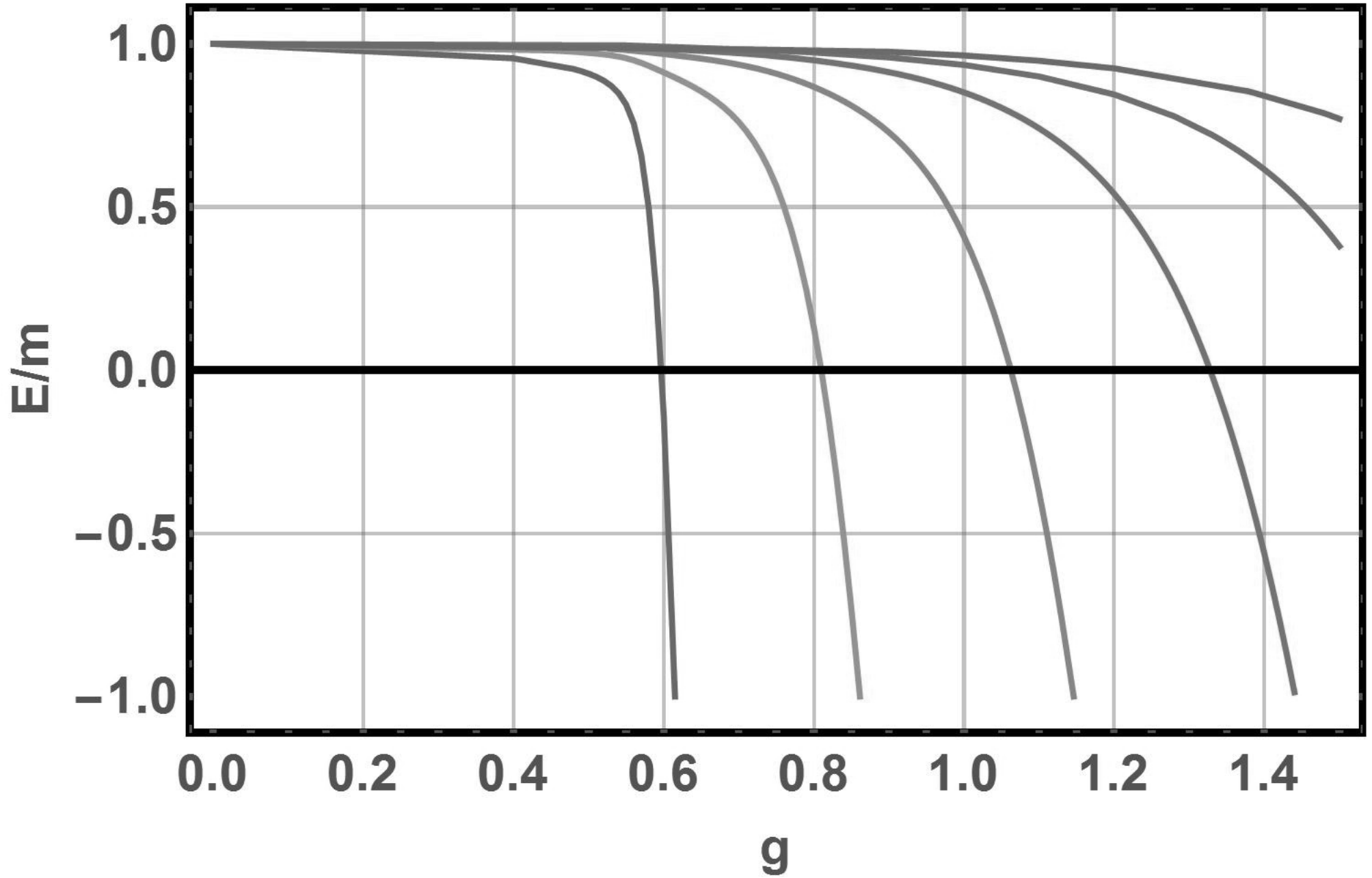}
}
\caption{The $g$ dependence of the lowest energy levels for $j=1/2$, $s=1$.}
\end{figure}

\section{Eigenstates for $E=\pm m$\label{R2}}

The preceding consideration fails for the points $E=m$ and $E=-m$ because
according to Eqs. (\ref{r9a}) and (\ref{r10}) at these points the variable
$\beta =\sqrt{m^{2}-E^{2}}$ becomes zero, while the factors $1/\sqrt{m-E}$,
$1/\sqrt{m+E}$, $b_{-}=\kappa +gm/\beta$, and the parameter
$a=\mu - gE/\beta$ blow up. Therefore, these points have to be considered
separately.

\subsection{The point \thinspace $E=m$\label{R2.2}}

It suffices to examine the set (\ref{r1}) and (\ref{r4}) with $E=m$ in the
exterior region $\rho _{0}\leq \rho <\infty$ where it has the form
\begin{align}
& f^{\prime }(\rho )+\frac{\kappa }{\rho }f(\rho )-
\left( 2m+\frac{g}{\rho }\right) g(\rho )=0,  \notag \\
& g^{\prime }(\rho )-\frac{\kappa }{\rho }g(\rho )+\frac{g}{\rho }f(\rho )=0.
\label{r20.1a}
\end{align}
It has to be supplemented by conditions that the both $f(\rho )$ and $g(\rho)$
functions are absolutely continuous and square integrable together with
their derivatives on $(\rho _{0},\infty)$.

Using second equation (\ref{r20.1a}), we find that the function $g(\rho)$
satisfies the second-order differential equation
\begin{equation}
g^{\prime \prime }(\rho )+\frac{1}{\rho }g^{\prime }(\rho )+
\left( \frac{2gm}{\rho }-\frac{\kappa ^{2}-g^{2}}{\rho ^{2}}\right) g(\rho )=0.
\label{r20.2a}
\end{equation}
The substitution $g(\rho )=w(z)$, $z=2\sqrt{2gm\rho }$ reduces
Eq. (\ref{r20.2a}) to the Bessel equation,
\begin{equation*}
w^{\prime \prime }(z)+\frac{1}{z}w^{\prime }(z)+
\left( 1-\frac{\tilde{\nu}^{2}}{z^{2}}\right) w(z)=0,
\end{equation*}
with $\tilde{\nu}=2\mu =2\sqrt{j^{2}-g^{2}}$,
$2\sqrt{2gm\rho _{0}}\leq z<\infty$. The general solution of this equation
is given by $w(z)=c_{1}H_{\tilde{\nu}}^{(1)}(z)+c_{2}H_{\tilde{\nu}}^{(2)}(z)$, where
$H_{\nu}^{(1)}(z)$ and $H_{\nu }^{(2)}(z)$ are the respective first and second
Hankel functions, see \cite{BatEr}. Its asymptotic behavior at infinity is
given by ($z\rightarrow \infty$)
\begin{eqnarray*}
w(z)&&=c_{1}\sqrt{\frac{2}{\pi z}}\exp
\left[ \frac{i}{4}\left( 4z-2\pi \tilde{\nu}-\pi \right) \right]
\left[ 1+O\left( \frac{1}{z}\right) \right] \\
&&\quad+c_{2}\sqrt{\frac{2}{\pi z}}\exp
\left[ -\frac{i}{4}\left( 4z-2\pi \tilde{\nu}-\pi \right) \right]
\left[ 1+O\left( \frac{1}{z}\right) \right].
\end{eqnarray*}
In view of second equation (\ref{r20.1a}), it follows that the asymptotic
behavior of both functions $f(\rho)$ and $g(\rho)$ at infinity is
estimated as $f(\rho)$, $g(\rho)=O(\rho ^{-1/4})$, $\rho \rightarrow \infty$,
so that the both functions are not square-integrable
at infinity. This means that system (\ref{r20.1a}) has no square-integrable
solutions and therefore, there are no bound states with the energy $E=m$,
i.e., with zero binding energy, for an electron in the Coulomb field of any
charge $Z$ with cutoff (\ref{r4}), as well as in the Coulomb field of a
point charge.

\subsection{The point \thinspace $E=-m$\label{R2.3}}

The system (\ref{r1}) of radial equations for bound states with $E=-m$, i.e.
with binding energy $2m$, becomes
\begin{align}
& f^{\prime}(\rho)+\frac{\kappa}{\rho}f(\rho)+V(\rho)g(\rho)=0,  \notag \\
& g^{\prime}(\rho)-\frac{\kappa}{\rho}g(\rho)-\left[ V(\rho)+2m\right]
f(\rho)=0,  \label{r21}
\end{align}
being supplemented by the conditions that the both $f(\rho)$ and $g(\rho)$
are absolutely continuous together with their first derivatives and square
integrable on $(0,\infty)$ and become zero at the origin, $f(0)=0$, $g(0)=0$.

\subsubsection{Solutions in the region $0\leq \protect\rho \leq \protect\rho_{0}$}

The general solution of system (\ref{r21}) and (\ref{r4}) in the interior
region $0\leq \rho \leq \rho_{0}$, where we set $f(\rho)=f_{in}(\rho)$
and $g(\rho)=g_{in}(\rho)$, under the above conditions is given by
\begin{align*}
f_{\mathrm{in}}(\rho )&=c\sqrt{\rho }J_{\nu }(\eta _{0}\rho )=c\sqrt{\rho}
\left\{
\begin{array}{l}
J_{|j|+1/2}(\eta _{0}\rho ),\ \mathrm{\ }\zeta =1, \\
J_{|j|-1/2}(\eta _{0}\rho ),\mathrm{\ }\zeta =-1,
\end{array}
\right. \\
g_{\mathrm{in}}(r)&=c\sqrt{\rho }
\sqrt{1-\frac{2m\rho _{0}}{g}}\zeta J_{\nu-\zeta }(\eta _{0}\rho ) 
=c\sqrt{\rho }\sqrt{1-\frac{2m\rho _{0}}{g}}\left\{
\begin{array}{l}
J_{|j|-1/2}(\eta _{0}\rho ),\mathrm{\ }\zeta =1, \\
-J_{|j|+1/2}(\eta _{0}\rho ),\text{\textrm{\ }}\zeta =-1,
\end{array}
\right.
\end{align*}
where $\eta_{0}=(g/\rho_{0})\sqrt{1-2m\rho_{0}/g}$.
It suffices to set $E=-m$ in (\ref{r5b}), (\ref{r7a}), and (\ref{r8a}).

\subsubsection{Solutions in the region $\protect\rho _{0}\leq \protect\rho <\infty $}

In the exterior region $\rho _{0}\leq \rho <\infty$, where we set
$f(\rho)=f_{out}(\rho)$ and $g(\rho )=g_{out}(\rho)$, Eqs. (\ref{r21}) and
(\ref{r4}) become
\begin{align}
& f_{out}^{\prime }(\rho )+\frac{\kappa }{\rho }f_{out}(\rho )-
\frac{g}{\rho}g_{out}(\rho )=0,  \notag \\
& g_{out}^{\prime }(\rho )-\frac{\kappa }{\rho }g_{out}(\rho )-
\left( 2m-\frac{g}{\rho }\right) f_{out}(\rho )=0,  \label{r24a}
\end{align}%
under conditions that both functions $f_{out}(\rho)$ and $g_{out}(\rho)$
are absolutely continuous and square-integrable together with their
derivatives on $(\rho_{0},\infty)$.

Using first Eq. (\ref{r24a}) in the second one, we find
\begin{eqnarray}
f_{out}^{\prime \prime }(\rho )+\frac{1}{\rho }~f_{out}^{\prime }(\rho )
-\frac{2gm}{\rho }f_{out}(\rho )-\frac{\kappa ^{2}-g^{2}}{\rho ^{2}}f_{out}(\rho )=0.
\label{r26a}
\end{eqnarray}
The substitution $f_{out}(\rho)=w(z)$, $z=2\sqrt{2gm\rho}$, reduces Eq. (\ref{r26a})
to the one for the modified Bessel functions, see \cite{BatEr},
\begin{eqnarray*}
w^{\prime \prime }(z)+\frac{1}{z}w^{\prime }(z)-
\left( 1+\frac{\tilde{\nu}^{2}}{z^{2}}\right) w(z)=0,\quad
\tilde{\nu}=2\mu = 2\sqrt{j^{2}-g^{2}},\quad
2\sqrt{2gm\rho _{0}}\leq z<\infty.
\end{eqnarray*}
\ The requirement for $f_{out}(\rho )$ to be square-integrable at infinity
then yields $f_{out}(\rho )=AK_{\tilde{\nu}}(z)$, where $K_{\tilde{\nu}}(z)$
is the McDonald function,
\begin{align*}
& K_{\tilde{\nu}}(z)=\frac{\pi }{2\sin \pi \tilde{\nu}}
\left[ I_{-\tilde{\nu}}(z)-I_{\tilde{\nu}}(z)\right] ,\
\ \tilde{\nu}\neq n\in \mathbb{Z}_{+}, \\
& I_{\tilde{\nu}}(z)=\sum\limits_{m=0}^{\infty }
\frac{(z/2)^{2m+\tilde{\nu}}}{m!\Gamma (m+\tilde{\nu}+1)},\
\ K_{\tilde{\nu}}(z)=K_{-\tilde{\nu}}(z).
\end{align*}
For $\tilde{\nu}=n\in \mathbb{Z}_{+}$, the functions $K_{n}(z)$ contain
terms with a logarithmic factor, see \cite{BatEr}.

We finally obtain a general solution of Eqs. (\ref{r24a}):
\begin{align*}
f_{out}(\rho )&=AK_{\tilde{\nu}}(z),\ \ z=2\sqrt{2gm\rho }, \\
g_{out}(\rho )&=\frac{A}{g}\left[ \frac{z}{2}\frac{d}{dz}K_{\tilde{\nu}}(z)+
\kappa K_{\tilde{\nu}}(z)\right] =\frac{A}{g}\left\{ -\frac{z}{4}\left[
K_{\tilde{\nu}-1}(z)+K_{\tilde{\nu}+1}(z)\right] +
\kappa K_{\tilde{\nu}}(z)\right\},
\end{align*}
where we have used the known formula $K_{\tilde{\nu}-1}(z)+
K_{\tilde{\nu}+1}(z)=-2K_{\tilde{\nu}}^{\prime }(z)$ (see \cite{BatEr}).

\section{Supercritical charges\label{R3}}

After the general solution of system (\ref{r21}) is found independently in
the respective regions $0\leq \rho \leq \rho _{0}$ and
$\rho_{0}\leq \rho <\infty $, it remains to satisfy the basic continuity condition
for the solution as a whole (to sew the partial solutions together smoothly), which
reduces to the requirement of continuity of the solution at the point $\rho=\rho_{0}$,
\begin{equation}
f_{in}(\rho _{0})=f_{out}(\rho _{0}),\
\ g_{in}(\rho _{0})=g_{out}(\rho_{0}).  \label{r29}
\end{equation}
The compatibility of equalities (\ref{r29}) with $c\neq 0$, $A\neq 0$ yields
the relation
\begin{gather}
J_{\nu }(\eta _{0}\rho _{0})\left\{ -\frac{z_{0}}{4}\left[
K_{\tilde{\nu}-1}(z_{0})+K_{\tilde{\nu}+1}(z_{0})\right] +\kappa K_{\tilde{\nu}}(z_{0})
\right\}
-(\eta _{0}\rho _{0})\zeta J_{\nu -\zeta }(\eta _{0}\rho_{0})K_{\tilde{\nu}}(z_{0})=0,  \notag \\
\eta_{0}\rho _{0}=g\sqrt{1-\frac{2m\rho _{0}}{g}},\quad z_{0}=2\sqrt{2gm\rho_{0}}, \label{r30a}
\end{gather}
that can be considered as an equation for coupling constants $g$ that
provides bound states with the energy $E=-m$. We let $g^{(-m)}(j,s)$ denote
such coupling constants.

An analytical solution of equation (\ref{r30a}) for $g^{(-m)}(j,s)$ with
arbitrary $j$, $s$ is unlikely to be possible at present. We only can try to
analyze it qualitatively and solve it numerically.

An equivalent form of equation (\ref{r30a}) that is more suitable for its
qualitative analysis and its numerical solution reads
\begin{eqnarray}
(\eta _{0}\rho _{0})\frac{\zeta J_{\nu -\zeta }(\eta _{0}\rho _{0})}
{J_{\nu}(\eta _{0}\rho _{0})}+\left[ \frac{z_{0}}{4}\frac{K_{\tilde{\nu}-1}(z_{0})+
K_{\tilde{\nu}+1}(z_{0})}{K_{\tilde{\nu}}(z_{0})}-\kappa \right]=0.
\label{r31a}
\end{eqnarray}
There exists an infinitely growing sequence
$\{g_{n}^{(-m)}(j,s)$, $n\in \mathbb{N}\}$,
$g_{n}^{(-m)}(j,s)\rightarrow \infty$ as
$n\rightarrow\infty$, of roots of this equation with any fixed
$j$, $s$, \cite{VorGitLevFer2016}. This infinite sequence yields the corresponding
infinitely growing sequence
$\{Z_{n}^{(-m)}(j,s)=\alpha_{F}^{-1}\epsilon ~g_{n}^{(-m)}(j,s)$,
$n\in \mathbb{N}\}$ of charges $Z$.
\begin{figure}[h]
\label{critical_charge}
\center{
\includegraphics[width=9cm]{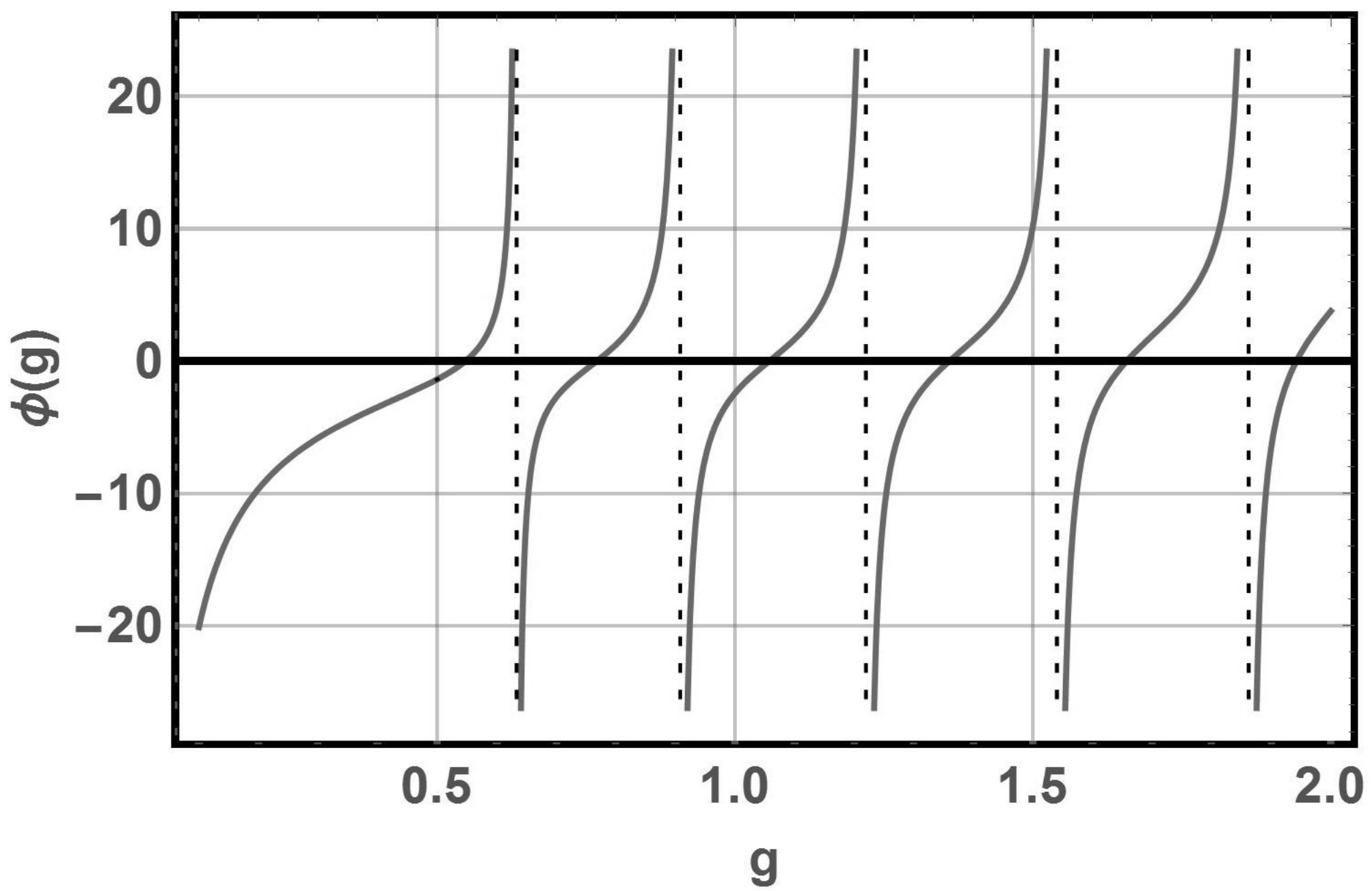}
}
\caption{Graphic solution of Eq. (\ref{r31a}).}
\end{figure}

We now define the supercritical charge $Z_{\mathrm{scr}}(j,s)$ for each pair
of quantum numbers $j$, $s$ as an integer nearest to $Z_{1}(j,s)$ (the first
term in the sequence $\{Z_{n}^{(-m)}(j,s)$, $n\in \mathbb{N}\}$) from above,
\begin{eqnarray*}
&&Z_{\mathrm{scr}}(j,s)=\left\{
\begin{array}{l}
\lbrack Z_{1}(j,s)]+1,\text{ }Z_{1}(j,s)\notin\mathbb{N},\\
Z_{1}(j,s),\text{ }Z_{1}(j,s)\in\mathbb{N},
\end{array}\right.\quad
Z_{1}(j,s)=(\alpha _{F}^{-1}\epsilon )g_{1}^{(-m)}(j,s),
\end{eqnarray*}
the symbol $[\dots]$ denotes the integral part of a real number.

We believe that the supercritical charge is defined by the minimum of all
the charges $Z_{n}^{(-m)}(j,s)$, which is attained in the sector $j=1/2$,
$s=1$ and is equal to
\begin{equation*}
Z_{\mathrm{scr}}=\left\{
\begin{array}{l}
\lbrack Z_{1}(1/2,1)]+1,\text{\textrm{\ }}Z_{1}(1/2,1)\notin\mathbb{N},\\
Z_{1}(1/2,1),\text{ }Z_{1}(1/2,1)\in\mathbb{N}.
\end{array}
\right.
\end{equation*}
The corresponding coupling $g_{\mathrm{scr}}=g_{1}^{(-m)}(1/2,1)$ also can
be called the supercritical coupling.

It is worth noting that the supercritical charges depend on a regularization
of the Coulomb field, in other words, the supercritical charges are model
dependent, and in addition, they depend on the parameters $\alpha _{F}$ and
$\epsilon$.

To determine $g_{n}^{(-m)}(1/2,1)$, it is convenient to represent Eq. (\ref{r31a})
as $\phi (g)=0$,
\begin{align*}
\phi (g)=(\eta _{0}\rho _{0})\frac{J_{1}(\eta _{0}\rho _{0})}
{J_{0}(\eta _{0}\rho _{0})}
-\left[ \frac{z_{0}}{4}\frac{\left[ K_{\tilde{\nu}-1}(z_{0})+
K_{\tilde{\nu}+1}(z_{0})\right]}{K_{\tilde{\nu}}(z_{0})}+\frac{1}{2}\right],
\end{align*}
where $\tilde{\nu}=\sqrt{1-4g^{2}}$.

Results of numerical calculations are presented graphically in Fig. 9, the first lower values of $g_{n}^{(-m)}(1/2,1)$ are:
\begin{eqnarray}
\{g_{n}^{(-m)}(1/2,1)\}=\big{\{}0.54731,\text{ }0.767737,\text{ }1.05737,
1.35964,\text{ }1.65711,...\big{\}},  \label{r34a}
\end{eqnarray}
for $m=0.26\,\mathrm{eV}$.
Sequence (\ref{r34a}) corresponds to the $\{Z_{n}^{(-m)}(1/2,1)\}$. Then, we
have $g_{\mathrm{scr}}=g_{1}^{(-m)}(1/2,1)=0.54731$.

\section{Conclusion\label{S5.4d}}

Solving spectral problems for electronic excitations in a graphene in the
presence of point-like Coulomb impurities, we have demonstrated that from
the mathematical standpoint, there is no problem in defining s.a.
Hamiltonians that define energy spectra and the corresponding complete sets
of eigenfunctions for any charge of the impurities. We have constructed
families of such possible s.a. Hamiltonians that are parameterized by some
extension parameters. The general theory thus describes all the
possibilities that can be offered to a physicist for his choice. This choice
is a completely physical problem.

Energy levels were calculated and corresponding (generalized) eigenfunctions
were obtained for any charge of the impurities
(see Eqs. (\ref{5.3.2})--(\ref{5.3.6})). Namely, for the nonsingular region
($g\leq g_{\mathrm{s}}(j)$) energy levels and normalized (generalized) eigenfunctions were
described by Eqs. (\ref{5.1a}) and (\ref{5.2a}), (\ref{5.2b}) respectively.
The same was done for the subcritical region
($g_{\mathrm{s}}(j)<g<g_{\mathrm{c}}(j)$), for the critical region
($g=g_{\mathrm{c}}(j)$), and for the overcritical region ($g>g_{\mathrm{c}}(j)$),
see Eqs. (\ref{4.1.2})--(\ref{4.1.2b}), (\ref{4.13b})--(\ref{4.1.3c}) and
(\ref{5.2.3.1b})--(\ref{5.2.3.1c}) respectively.

We stress that the obtained eigenfunctions can be used to calculate a local
density of states, which can be measured experimentally by using the
scanning tunneling microscopy \cite{Kat}. The importance of calculations in
the graphene with Coulomb impurities is confirmed by results of the work
\cite{Zhu}, where it was shown that, in contrast to an undoped graphene,
there are significant differences in the behavior of a local density of
states near the boundary of the positive continuum.

It is interesting to note that in our problem, the critical coupling
$g_{\mathrm{c}}(\pm 1/2)=1/2$ and the lower critical coupling
$g_{\mathrm{s}}(\pm 1/2)=0$, whereas, in the $3$-dimensional case they are
$g_{\mathrm{c}}(1/2)=1$, $g_{\mathrm{s}}(1/2)=\sqrt{3}/2$. Besides, due to the
large value of the fine structure constant in graphene, $\alpha _{F}\simeq 2.2$, the
critical value of the impurity charge is small:
$Z_{\mathrm{c}}=Z_{\mathrm{c}}(\pm 1/2)\simeq 1$, which opens the
possibility of testing the supercritical instability in the graphene \cite{Kotov2012}.
Indeed, the supercritical atomic collapse in graphene was observed experimentally and
reported in Ref. \cite{ExpGc}.

In contrast to the $3$-dimensional case, in the problem under consideration
for all values of impurity charge corresponding s.a. Hamiltonians are not
defined uniquely. We recall that the Dirac Hamiltonian for an electron in
the Coulomb field in $3$ dimensions is defined uniquely for $Z\leq 118$. We
note that a transition through critical charges does not lead to any
technical qualitative changes in the mathematical description.

It should be noted that in the Ref. \cite{Gupta} a bound-state spectrum of
low-energy excitations in a gapped graphene with a charged impurity, was studied
without a convinced analysis of the asymptotic behavior of wave functions based
on a correct construction of a corresponding s.a. Dirac Hamiltonian. 
Besides, in the Ref. \cite{Kh2011} s.a. Dirac Hamiltonians
with the Coulomb field in combination with the Aharonov-Bohm field in $2+1$
dimensions and their spectral analysis were considered. However, in this
consideration, features of the graphene problem were not taken into account.
Because of this, the radial Hamiltonians which were considered in this work,
are parametrized in a specific way which does not allow one to identify them
with the corresponding Hamiltonians of real graphene problem. Moreover, the
zero limit of the additional external field (the Aharonov-Bohm field), which
is necessary for possible comparison, was not studied and seems like is a
nontrivial task.

Studying the spectral problem in the graphene with impurities presented by a
regularized Coulomb potential (see Section \ref{r}), we have found the
so-called supercritical charges, for which lowest levels of energy spectra
reach the value $-m$. Formally (see discussion in Ref. \cite{VorGitLevFer2016},
where the corresponding $3+1$ dimensional case was studied, and Refs. \cite{CrD,ZelPo72}),
this may be an indication that for such charges the vacuum becomes
unstable with respect to possible pair creation. At the same time this may be
an indication that the problem becomes many particle one, such that one-particle
relativistic quantum mechanics based on the Dirac Hamiltonian fails. Then, calculations
in the framework of the latter model may be not sufficient for statements about the
existence of real physical effects such as particle production. For this
reason, we believe that the problem of the production of electron-positron
pairs from the vacuum by a supercritical Coulomb field is still far from its
completion. That is why we cannot accept the conclusion of the work \cite{Kyl} that a
real production of electron-positron pairs by a regularized supercritical Coulomb field
is impossible.

Note that the Eq. (\ref{r31a}) that defines the supercritical coupling
constant $g_{\mathrm{scr}}$ has solutions only for
$g_{\mathrm{scr}}>g_{\mathrm{c}}(\pm 1/2)=0.5$. For $\rho_{0}=0.6a$ and $m=0.26\,\mathrm{eV}$,
we obtain $g_{\mathrm{scr}}=0.54731$, which corresponds to small values
$Z_{\mathrm{scr}}\simeq Z_{\mathrm{c}}\simeq 1$ for the dielectric constant in
the range $\epsilon \approx 2.4-5$. Thus, even after the regularization, the
supercritical charge is the same as critical charge for the point-like
Coulomb field case. We note that in $3$-dimensional case, for the
regularized Coulomb potential the supercritical charge $Z_{\mathrm{scr}}=174$,
is greater than the corresponding critical charge $Z_{\mathrm{c}}=138$
\cite{VorGitLevFer2016}.

Finally, it should be noted that in Refs. \cite{Godb, Godb2}, they have used
the same regularization for the Coulomb field of impurities, then an
equation for the spectrum (up to the notation) in the case of $s=1$ had the
form (\ref{r17}). However, results of numerical calculations refer to
different physical parameters than in our consideration. Moreover,
calculations of a critical charge were done only in the zero limit of the
cutoff parameter. We note that a consideration of critical charges, similar
to the present work, was undertaken in Refs. \cite{Pereira2} and \cite{Zhu},
where, however, the discrete spectrum was not investigated.

\section{Acknowledgements}

The work is supported by Russian Science Foundation (Grant No. 19-12-00042).

\end{document}